# Dislocation Dynamics in a Crystal Lattice (Peierls-Nabarro) Relief

B. V. Petukhov

Institute of Crystallography, Russian Academy of Sciences, Moscow, Russia

**Abstract**

The theory of the dislocation motion in the periodic potential relief of the crystal lattice (the Peierls-Nabarro barriers) is reviewed. On the basis of the kink mechanism the temperature dependence of the flow stress is described for a wide class of materials. The theory of quantum mechanical dislocation tunnelling through the Peierls-Nabarro barriers is extended and compared with experimental data on the plasticity of alkali halides, BCC and HCP metals at low temperatures. The behavior of the flow stress at the range of athermic anomalies is modeled by changing the mechanism of the dislocation motion from the thermally activated hopping over the barriers to the quantum tunnelling through them. Some results of previous calculations are represented in a more explicit convenient for applications form. The pronounced effect of the switching between the normal and the superconducting states on the flow stress of metals is explained on the basis of the change in the dissipative properties of the electron subsystem affecting the dislocation motion.

## Contents

## 1. INTRODUCTION

Extended defects of topological character, such as dislocations, are specific among different types of crystal defects, due to their ability to move under the action of relatively small forces. Many important mechanical and electrophysical properties of materials are determined by dislocation dynamics, such as plasticity (e. g., Hirth and Lothe 1982, Nadgornyi 1988, Suzuki, Takeuchi and Yonenaga 1991), internal friction (Seeger and Schiller 1966), degradation of electronic devices (Hull, Bean and Logan 1993), etc. The mobility of dislocations crucially depends on the pinning mechanisms. Dislocations may be pinned by imperfections of the crystal structure (impurities, precipitates, etc). However, even in the sufficiently perfect crystal materials there exist intrinsic barriers for dislocation motion, caused by the discreteness of their crystal lattices. These barriers (called as Peierls-Nabarro barriers) in agreement with translation symmetry of the crystal are periodic, with lattice parameter $a$. Their magnitude may be characterized by stress of the pinning, or Peierls stress $\sigma_P$, giving maximum to the resistance force $\sigma_P b$ exerted by the crystal lattice, per unit length of the dislocation line. Here $b$ is the Burgers vector of dislocation, or its topological charge. Since dislocation mobility is affected by the Peierls-Nabarro barriers, the mechanical properties of materials depend on such fundamental characteristics as the type of crystal structure and character of chemical bonds between the atoms. The value of $\sigma_P$ varies between $10^{-5}$-$10^{-4}\mu$ ($\mu$ is the shear modulus) for FCC and HCP materials, $10^{-4}$-$10^{-3}\mu$ for BCC metals, $10^{-3}$-$10^{-1}\mu$ for ceramics and semiconductors.

The most popular form of potential used for quantitative description of the Peierls-Nabarro relief is harmonic potential

$$U_0(y) = \frac{\sigma_P ab}{2\pi}[1 - \cos(\frac{2\pi y}{a})], \tag{1.1}$$

where $y$ is dislocation displacement in a slip plane. Sometimes, for a better description of experimental data, a number of additional harmonics are included in the potential (1.1) (Guyot and Dorn 1967). Note that $U_0(y)$, in accordance with extended character of dislocation, is the energy per unit length.

The state of a dislocation in the valleys of Peierls-Nabarro relief is periodically degenerated, however states in the neighbouring valleys are separated by a barrier, so that transitions between the valleys may occur due to fluctuations providing the required energy. In the presence of an external stress $\sigma$ the potential (1.1) is transformed into the so-called "wash-board" inclined potential shown in Fig. 1:

$$U(y) = U_0(y) - \sigma b y, \tag{1.2}$$

External loading eliminates the degeneration and makes the dislocation state in any valley metastable, with a limited life-time. The dislocation starts moving, fluctuationally, towards more energetically favourable states.

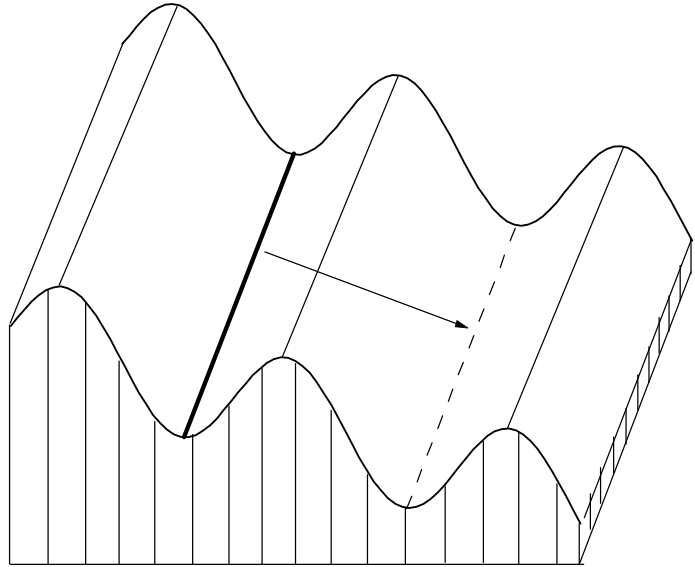

Fig. 1. The potential $U(y)=U_0(y)-\sigma by$ for the dislocation displacement $y$ in the slip plain in the presence of an external stress (the washboard potential). The dash line marks the bottom of the Peierls-Nabarro relief valley, the bold line shows a dislocation position.

The decay of metastable states of an extended system like a dislocation, having macroscopic number of the freedom degrees, proceeds similarly to the first order phase transition, namely in two-steps. At first, fluctuations throw a small portion of dislocation over the barrier into the next valley of relief having lower energy level. Thus a nucleus for the transition is formed. Then the kinks surrounding the nucleus slip under the action of external load, along the barrier ridge, up to the ends of dislocation segment, or until merging with the neighbouring kink pair. This completes the transition of dislocation into the next valley, representing thus an elementary step of dislocation motion over the lattice period. The dislocation advance may be represented as formation of kink pairs and their sideways motion. Kinks, as specific quasiparticles, provide a convenient alternative model for description of dislocation dynamics. In the same way as dislocations describe the "soft mode" of crystal deformation, the kinks characterize the soft mode of dislocation motion on the elementary level.

By analogy with Peierls-Nabarro barriers for dislocation motion in the slip plane, one may expect that periodical potential relief exists for the kink migration along the dislocation line, the so-called Peierls-Nabarro barriers of second kind. This barriers can be significant in materials like semiconductors, where the kink width $d$ is small and comparable with the lattice period $a$. However, in most of metals and many other materials due to high value of the dislocation line energy $\sim\mu b^2$ (versus Peierls-Nabarro potential), the kink width, estimated as $d\sim a(\mu/\sigma_P)^{1/2}$, is large, so that secondary relief is smoothed out.

Clearly, due to topological limitation, kinks are created and annihilated in crystal only by pairs. Contrary to phonons, they are strongly non-linear excitations, often referred to as topological solitons, or kink-solitons (see e.g., Dodd 1984). From historical viewpoint, it is interesting to note that the first application of soliton concepts to description of experiments in solid state physics was apparently related with dislocation kinks (Mott and Nabarro 1947).

Since the kinks slip easily in materials with low secondary relief, the "bottle-neck" of dislocation transition between two neighbouring valleys is actually a process of kink pair formation. If the average transition time $t_{tr}$ is known, one can easily find the over-all dislocation velocity $V$ as $V=a/t_{tr}$. In simple cases this allows also to describe the macroscopic plastic flow of materials $\dot{\varepsilon}$, making use of the Orowan relationship (e. g., Suzuki, Takeuchi and Yoshinaga 1991),

$$\dot{\varepsilon} = \rho b V, \qquad (1.3)$$

where $\rho$ is the density of mobile dislocations. Obtaining of the stress and temperature dependence of transition time is the main objective of this paper.

## 2. POTENTIAL RELIEF FOR DISLOCATION MOTION

Kink-pair formation assumes a considerable deviation of the dislocation line from a straight position along the bottom of Peierls-Nabarro relief valley. Such nonuniform variation of the dislocation configuration additionally modifies the dislocation energy. To describe the energy variation with a change of dislocation shape, the uniform one-dimensional Peierls-Nabarro potential is insufficient. One has to use a generalized multidimensional potential in the space of possible dislocation configurations. The simplest expression for such potential relief is given in the line tension approximation (e. g., Suzuki, Takeuchi and Yoshinaga 1991)

$$E\{y(x)\} = \int [\frac{\kappa}{2}(\frac{dy}{dx})^2 + U(y(x))]dx . \qquad (2.1)$$

Here $y(x)$ is a nonuniform dislocation configuration ($x$ is the coordinate along the bottom of the valley), $k$ is the line tension constant. The additional term in equation (2.1) describes an elongation of dislocation due to its bending, and the corresponding growth of its energy. Clearly, such simple approximation is applicable only for smooth dislocation configurations with $dy/dx<<1$.

For a rectilinear dislocation $y=const$, and the equation (2.1) is reduced to $E\{y\}=U(y)L$ where $L$ is the length of considered dislocation segment. Thus, (2.1) yields the same value per unit length as the equation (1.2). Also, overcoming of the barrier by a long dislocation segment as a whole is seen to require too much activation energy, so that less expensive may be local transition with deviation from rectilinear shape. In order to find this way and calculate the encountered barriers, we shall first investigate the "topography" of potential relief (2.1).

A simplified schematic 2D-illustration of this multidimensional relief is shown in Fig. 2. Minima of the relief (2.1) are separated by "ridges", the lowest of which, $E_0$, corresponds to a "saddle point" $y_0(x)$, a dislocation configuration being a nucleus for a subsequent formation of a kink pair. ($E_0$ is activation energy of the process.)

Saddle configuration $y_0(x)$ can be found from extremum of the functional (2.1), i.e. from equation

$$\frac{\delta E}{\delta y(x)} = -\kappa \frac{d^2 y_0}{dx^2} + \frac{dU(y_0(x))}{dy} = 0 . \qquad (2.2)$$

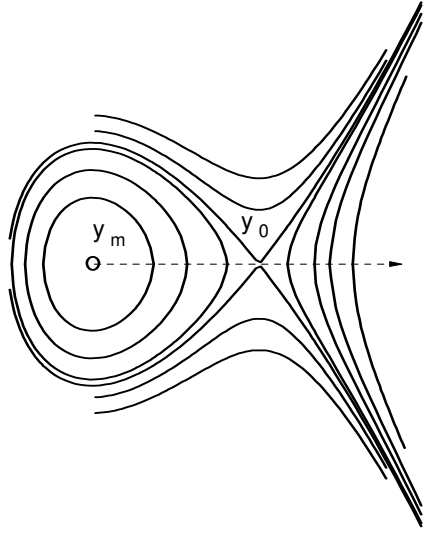

Fig. 2. A schematic illustration of the multidimensional potential relief in the vicinity of the saddle point ($y_0$). Lines correspond to a constant value of the energy, $y_m$ is minimum of the potential.

This equation is similar to the Newton equation for mechanical motion of a particle in the potential $-U(x)$ (where $\kappa$ and $x$ play role of the mass and the time). The first integral of (2.2), which in mechanical analogy means the energy conservation law, can be written as

$$\frac{\kappa}{2}\left(\frac{dy}{dx}\right)^2 - U(y) = const. \quad (2.3)$$

The constant can be determined from the boundary condition at $x \to \infty$: $y_0(x) \to y_m$ ($y_m$ is the minimum of $U(x)$), $dy_0/dx \to 0$. Therefore, $const.=-U(y_m)$, and we have for $y_0(x)$ from the equation (2.3)

$$x = \pm\sqrt{\frac{\kappa}{2}}\int_{y_m}^{y_0}\frac{dy}{\sqrt{U(y)-U(y_m)}} \,. \quad (2.4)$$

This relation solves the problem of finding the saddle configuration $y_0(x)$. Substituting it into (2.1) yields the following expression for the activation energy $E_0(\sigma)$ (Celli *et al.* 1963).

$$E_0(\sigma) = 2\int_{y_m}^{y_M}\sqrt{2\kappa[U(y)-U(y_m)]}\,dy \,. \quad (2.5)$$

Here $y_M$ is the maximum of the dislocation displacement in the saddle configuration.
In the case of particular interest, when $\sigma \to 0$, $E_0(\sigma)$ is equal to doubled energy of an isolated kink $E_k$, $E_0(\sigma)=2E_k$, and (2.4) describes the shape of a kink. For the harmonic potential (1.1), this kink is the well-known sin-Gordon soliton (e. g., Dodd *et al.* 1984)

$$y^{\pm}(x)=(2a/\pi)\text{arctg}\exp\{\pm\sqrt{\frac{2\pi\sigma_P b}{\kappa a}}x\},\tag{2.6}$$

and its energy is $E_k=(2/\pi)^{3/2}(\kappa\sigma_P b a^3)^{1/2}$ (Fig. 3).

For a number of various potentials $U_0(y)$ the dependence of activation energy $E_0(\sigma)$ on stress in the whole range $0<\sigma<\sigma_P$ was calculated in the papers (Celli et al. 1963, Guyot and Dorn 1967). For harmonic Peierls-Nabarro potential (1.1), the result of $E_0(\sigma)$ calculation is plotted in Fig. 4. Analytical approximation in the wide stress range valid for most of applications is given by the simple expression (Baufeld *et al.* 1998)

$$E_0(\sigma)=2E_k[1-(\sigma/\sigma_P)^{0.8}]^{1.3}.\tag{2.7}$$

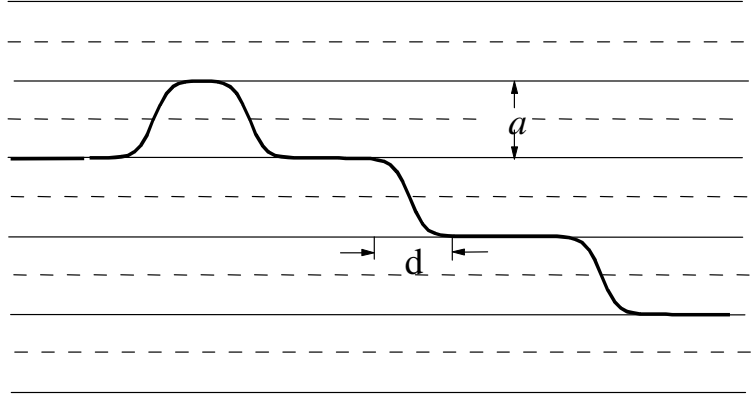

Fig. 3. The kinked dislocation in the Peierls-Nabarro relief. Solid lines are maxima (ridges), dashed lines are minima of the relief. *a* is the period of the relief, *d* is the kink width.

For particularly interesting cases of low ($\sigma\to 0$), and high ($\sigma\to\sigma_P$) stress values, more exact formulae may be obtained. In the following we shall pay a special attention to the stress range close to the Peierls stress $\sigma_P$, which is important for the low temperature dislocation dynamics. The detail characterization of the stress dependence of the activation energy in the range of the low stresses is given in (Suzuki, Koizumi and Kirchner 1995).

The carrying capacity of the barrier is determined mainly by the arrangement of the energetic relief in the vicinity of the saddle point. At small deviations from the saddle configuration $y(x)=y_0(x)+\delta y(x)$ the dislocation energy (2.1) can be approximately expressed as a quadratic form

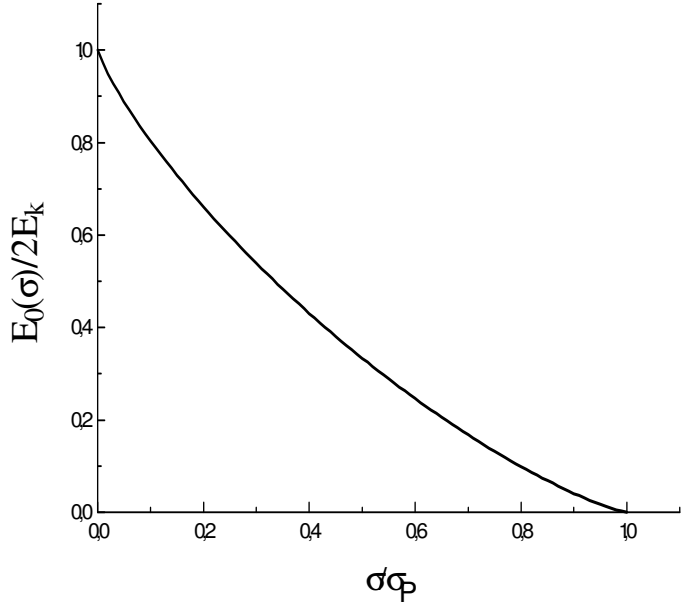

Fig. 4. The stress dependence of the kink pair formation energy $E_0(\sigma)$ for the harmonic Peierls-Nabarro potential.

$$E\{y\} \approx E_0 + \frac{1}{2}\int \frac{\delta^2 E}{\delta y(x)\delta y(x')}\delta y(x)\delta y(x')dxdx' = E_0 + \frac{1}{2}\sum_{\alpha=0}^{\infty}\mu_\alpha y_\alpha^2. \qquad (2.8)$$

Here $y_\alpha$ are projections of $\delta y(x)$ onto the main directions of the potential relief at the saddle point $y_0(x)$, $\mu_\alpha$ are the eigenvalues of the quadratic form, or the curvatures of the relief along the corresponding directions. Coordinate $y_0$ describes the pass over the saddle point, the corresponding eigenvalue $\mu_0$ is negative.

The equation for the eigenvectors $\phi_\alpha$ and eigenvalues $\mu_\alpha$ of the quadratic form (2.8) is

$$-\kappa d^2\phi_\alpha/dx^2 + U''(y_0(x))\phi_\alpha = \mu_\alpha \phi_\alpha. \qquad (2.9)$$

Let us note some general properties of the eigenvalues spectrum. There are a few of discrete eigenvalues, and a continuos spectrum, ranging between the asymptotic value $U''(y_0(x))$ at $x\to\pm\infty$, i. e., $U''(y_m)$, and infinity. The behaviour of the eigenvalue density $\rho(\mu)$ for $\mu\to\infty$ can be found by the perturbation theory

$$\rho(\mu)-\rho_0(\mu) \approx \frac{J}{4\pi\mu^{3/2}}, \qquad (2.10)$$

where $\rho_0(\mu)$ is the unperturbed density, describing fluctuations around the initial straight dislocation position along the "wash-board" potential $U(y)$ valley bottom, and

$$J=\sqrt{2}\int_{y_m}^{y_M}\frac{U''(y_m)-U''(y)}{\sqrt{U(y)}}dy. \qquad (2.11)$$

It is useful to have in mind, however, that equation (2.9), based on the line tension approximation, provides the correct description of the spectrum only if the corresponding space scale of the eigenfunctions $\phi_\alpha(x)$, which is of order of $(\kappa/\mu)^{1/2}$, is large in comparison with the lattice period $a$, i. e., $\mu \leq \kappa/a^2$.

Among the discrete eigenvalues there is a negative one, $\mu_0$, corresponding to the pass over the saddle point, and the zero eigenvalue $\mu_{tr}=0$, corresponding to the translation of the saddle configuration $y_0(x)$ along the Peierls-Nabarro relief valley, which does not change the energy of the dislocation.

For more complete description of the eigenvalues spectrum the Peierls-Nabarro potential has to be specified. A universal picture takes place for the stress range close to the Peierls stress $\sigma_P$, when any potential $U(y)$ in the vicinity of the merging of minimum and maximum can be approximated by the polynomial expression

$$U(y)=[\beta b(\sigma_P-\sigma)]^{1/2}y^2-\beta y^3/3, \qquad (2.12)$$

where $y$ and $U(y)$ are measured from the stable equilibrium (straight) configuration at the minimum of $U(y)$, and $\beta=|d^3U(y_P)/dy^3|/2$, $y_P$ is the inflection point of $U_P(y)$ (see Fig. 5). The approximation (2.11) is valid, when the distance between maximum and minimum of $U(y)$, $2[b(\sigma_P-\sigma)/\beta]^{1/2}$, is small in comparison with the relief period $a$. For typical Peierls-Nabarro potentials this condition fulfils when $\delta=(\sigma_P-\sigma)/\sigma_P<<1$. This case, which is of special interest for the low temperature dislocation dynamics, can provide a useful illustration of the general conclusions made above. One easily obtains from equation (2.4)

$$y_0(x)=\frac{3[b(\sigma_P-\sigma)/\beta]^{1/2}}{\cosh^2\{[\beta b(\sigma_P-\sigma)]^{1/4}x/(2\kappa)^{1/2}\}}. \qquad (2.13)$$

One may see that, whereas the amplitude of the nucleus $y_M=3[b(\sigma_P-\sigma)/\beta]^{1/2}$ decreases as $\sigma\to\sigma_P$, its width, or the size along the $x$ coordinate is of order $l_0=\kappa^{1/2}[\beta b(\sigma_P-\sigma)]^{-1/4}$, and it increases as $\sigma$ approaches $\sigma_P$.

The activation energy $E_0(\sigma)$ calculated from equation (2.5) is equal to

$$E_0(\sigma)=(24\times 2^{1/2}/5)\kappa^{1/2}[b(\sigma_P-\sigma)]^{5/4}/\beta^{3/4}. \qquad (2.14)$$

The solution of the eigenvalue problem (2.9) for this case is given, for example, in (Landau and Lifshitz 1958b). The discrete spectrum consists of three eigenvalues: $\mu_0=-(5/2)\mu_N$, $\mu_1=0$, and $\mu_2=(3/2)\mu_N$ ($\mu_N=[\beta b(\sigma_P-\sigma)]^{1/2}$). Besides, one can obtain the variation of the eigenvalues density
in respect to the undisturbed one

$$\rho(\mu)=\rho_0(\mu)-\frac{1}{\pi}\sqrt{\frac{\mu_N}{\mu-2\mu_N}}\left(\frac{1/2^{1/2}}{\mu-3\mu_N/2}+\frac{2^{1/2}}{\mu}+\frac{3/2^{1/2}}{\mu+5\mu_N/2}\right). \qquad (2.15)$$

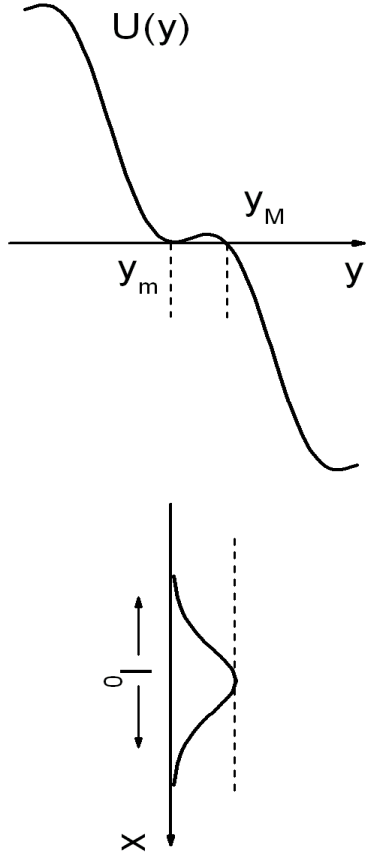

Fig. 5. The potential $U(x)$ and the kink pair nucleus shape $y_0(x)$ at stresses close to the Peierls stress $\sigma_P$. $l_0$ is the kink pair nucleus size.

For $\mu \to \infty$ this expression transforms into equation (2.10) with

$$J = 12(2)^{1/2}[\beta b(\sigma_P - \sigma)]^{1/2} \tag{2.16}$$

Now we have all necessary information about the arrangement of the potential relief for dislocations and may start to analyze the kinetics of processes proceeding in it.

## 3. KINK PAIR FORMATION RATE

Let us return to the general case $\sigma \sim \sigma_P$. The Arrhenius factor $\exp(-E_0(\sigma)/kT)$ provides the most essential temperature and stress dependencies of the kink pairs generation rate $\Gamma$. To obtain the total expression for $\Gamma$ we shall use the extended to multidimensional systems Kramers formula for the thermally activated process rate

$$\Gamma = \left(\frac{\Omega}{2\pi}\right)\left(\frac{Z_M}{Z_0}\right)\exp\left\{-\frac{E_0}{kT}\right\}. \tag{3.1}$$

Here $\Omega = (|\mu_0|/\rho + \eta^2/4\rho^2)^{1/2} - \eta/2\rho$, $\rho$ and $\eta$ are mass and viscosity (in our case per unit length of the dislocation), $Z_0$ is the partition function for fluctuations around the initial state at the minimum in front of the barrier, $Z_M$ is the partition function for fluctuations around the transitional state at the top of the barrier.

The partition functions $Z_0$ and $Z_M$ allow the factorization and can be represented as products of partition functions $Z_\alpha$ for each independent coordinate $y_\alpha$

$$Z_M \sim \int \exp\left[-\frac{\mu_\alpha y_\alpha^2}{2kT}\right] dy_\alpha = \left(\frac{2\pi kT}{\mu_\alpha}\right)^{1/2}. \tag{3.2}$$

The ratio $Z_M/Z_0$ can be expressed through the coefficients of the quadratic expansion of the energy (2.8) over fluctuations near the top of the barrier ($\mu_\alpha$), and near the initial minimum in front of the barrier ($\mu^0_\alpha$)

$$\frac{Z_M}{Z_0} = L\left(\frac{E_0}{2\pi\kappa kT}\right)^{1/2} \frac{\prod_{\alpha=0}^{\infty} \sqrt{\mu^0_\alpha}}{\prod_{\alpha=0}^{\infty}{}' \sqrt{|\mu_\alpha|}}. \tag{3.3}$$

The prime at the product sign in the denominator indicates that the eigenvalue corresponding to the translation mode $\mu_{tr}=0$ is omitted. A special consideration (Langer 1969, Petukhov and Pokrovskii 1973, Buttiker and Landauer 1981) shows that its contribution provides the factor, proportional to the dislocation segment length $L$, before the infinite products in (3.3). The length dependence in (3.3) has a simple meaning: it reflects the freedom of the choice of the kink pair birthplace along the dislocation line, which increases the statistical probability of a kink pair formation event. Therefore, $\Gamma$ may be represented as $\Gamma=gL$, where $g$ is the kink pair formation rate per unit length of the dislocation.

The difference between $\mu_\alpha$ and $\mu^0_\alpha$ vanishes when $\alpha$ increases, and, consequently, the main contribution to the ratio results from the first few factors. Since there is one more factors in the nominator, the order of magnitude of the ratio is $\sim |\mu_0|^{1/2} \sim \sigma_P^{1/2}$, and $Z_M/Z_0 \sim L(E_0\sigma_P/\kappa kT)^{1/2}$.

For the stress close to the Peierls stress the direct calculation of the products of the eigenvalues with the help of equation (2.15) yields (Petukhov (1986)

$$\Gamma = \frac{3 \cdot 2^{7/4}}{\pi^{3/2}} \left\{\left[\frac{\eta^2}{4\rho^2} + \frac{5}{2\rho}\sqrt{\beta b(\sigma_P - \sigma)}\right]^{1/2} - \frac{\eta}{2\rho}\right\} \frac{L[b(\sigma_P - \sigma)]^{7/8}}{\beta^{1/8}\kappa^{1/4}(kT)^{1/2}} \exp\left\{-\frac{E_0(\sigma)}{kT}\right\}. \tag{3.4}$$

The time for the dislocation shift $t_{tr}$ over one period of the crystal relief, when it is determined by the waiting time for the kink pair formation $1/\Gamma$, is inversely proportional to the dislocation segment length $L$. For sufficiently large $L$ the time of the kink pair expansion over the total dislocation segment $L/v_k$ ($v_k$ is the kink velocity) may be comparable to $1/\Gamma$, and the mechanism controlling the transition time is changed. Comparing the waiting time $1/gL$ and the expansion time $L/v_k$, one can determine the characteristic length $L_0=(v_k/g)^{1/2}$. For $L>L_0$ the length dependence of the transition time saturates, and one has (Lothe and Hirth 1959)

$$t_{tr}=1/gL, \qquad L<L_0, \tag{3.5}$$

$$t_{tr}=1/(gv_k)^{1/2}, \quad L>L_0. \tag{3.6}$$

The saturation of the transition time and the dislocation velocity length dependence is especially important in materials with a high secondary Peierls-Nabarro relief for the kink migration along the dislocation line, since the kink velocity $v_k$ is small.

## 4. TEMPERATURE DEPENDENCE OF FLOW STRESS

The dislocation motion mechanisms are investigated in experiments of different levels, from macroscopic mechanical tests, to measurements of the mobility of individual dislocations. Recently developed experimental technique for the investigation of the individual dislocation mobility in perfect semiconducting crystals under two-level intermittent loading allows for studying even modes of movement of the kinks along a dislocation line (Iunin et al. 1991). However, at present investigations of macroscopic plastic deformation of crystalline materials are practised on a wider scale.

The calculation of the kink pair formation rate $\Gamma$ and the dislocation velocity $V$ provides the basis for the quantitative description of the macroscopic strain rate $\dot{\varepsilon}$ by the Orowan relationship (1.3) $\dot{\varepsilon}=\rho_d bV$. The nucleation rate $\Gamma$ is given mainly by the Arrhenius factor $\exp(-E_0(\sigma)/kT)$, $E_0(\sigma)$ being the activation energy of equation (2.5). After all, one has

$$\ln(\frac{\dot{\varepsilon}_0}{\dot{\varepsilon}}) = \frac{E_0(\sigma)}{kT}. \tag{4.1}$$

For simplicity, the pre-exponential factor $\dot{\varepsilon}$ is considered as a temperature and stress independent constant. For experiments at a constant strain rate, the relationship (4.1) determines a temperature dependence of the flow stress $\sigma(T)$. As it follows from equation (4.1), $\sigma(T)$ is the inverted and re-scaled stress dependence of the activation energy. One more important consequence from equation (4.1) is the relationship

$$\ln\left(\frac{\dot{\varepsilon}_0}{\dot{\varepsilon}}\right)=-T\frac{\partial\sigma/\partial T}{\partial\sigma/\partial\ln\dot{\varepsilon}}. \tag{4.2}$$

This relationship predicts that the combination of the experimentally measured parameters on right hand of equation (4.2) should be a temperature independent constant for the mechanical tests at a constant strain rate $\dot{\varepsilon}$. Therefore, equation (4.2) may serve as a criterion of operation of the thermally activated mechanism of the plastic flow.

In the range of stresses close to the Peierls stress equations (2.14) and (4.1) yield

$$\delta\sigma=\sigma_P-\sigma(T)= A_1 T^{4/5}, \tag{4.3}$$

where $A_1=0.25(5/6)^{4/5}(\beta^{3/5}k^{4/5}/b\kappa^{2/5})\ln^{4/5}(\dot{\varepsilon}_0/\dot{\varepsilon})$. This type of the temperature dependence of the flow stress was experimentally verified for a wide class of materials (Gektina, Lavrent'ev and Natsik 1980, Kato 1983, Ackerman, Mughrabi, Seeger 1983, Brunner, Diehl and Seeger 1984, Haasen, Barthel and Suzuki 1984, etc.). Using it, one may simplify the operation criterion of the thermally activated kinetics (4.2). Its shortcoming is a considerable scattering of data on the temperature dependence of the flow stress, which leads to large errors in the estimation of the temperature sensitivity $\partial\sigma/\partial T$. This difficulty can be avoided, if for the smoothing data on $\partial\sigma/\partial T$ in the low temperature range one uses the relation $\partial\sigma/\partial T = -4\delta\sigma/5T$, following from equation (4.3), and allowing to express the criterion (4.2) for the stress range $\delta\sigma \ll \sigma_P$ through directly measured parameters

$$\ln\left(\frac{\dot{\varepsilon}_0}{\dot{\varepsilon}}\right) = \frac{4}{5}\frac{\delta\sigma}{\partial\sigma/\partial\ln\dot{\varepsilon}}. \tag{4.4}$$

The relationship (4.4) establishes also similarity between the temperature dependences of $\delta\sigma$ and the strain rate sensitivity $I = \partial\sigma/\partial\ln\dot{\varepsilon}$. The strain rate sensitivity, in turn, is directly related to a characteristic of the microscopic mechanism, the activation volume $\gamma = -dE_0/d\sigma$,

$$I = kT/\gamma. \tag{4.5}$$

## 5. INFLUENCE OF LOCAL OBSTACLES ON THE THERMOACTIVATED KINK PAIR FORMATION

Local crystal defects, like impurities, in many cases increase the flow stress (so-called "solid solution hardening effect" (e.g., Nadgorny 1988, Suzuki, Takeuchi and Yoshinaga 1991)). Let us consider a simple model of modification of the kink mechanism by the impurities. In the presence of a sufficient amount of impurities they divide the dislocation segment into small parts, and the kink pairs nucleated under these constraints. At low temperatures, when the flow stress approaches the Peierls stress, $\sigma \to \sigma_P$, the space scale of the kink pair increases proportionally to $(\sigma_P - \sigma)^{-1/4}$, and the space limitation due to the presence of impurities becomes more and more essential. As a result, the activation energy $E(\sigma, l)$ of the kink pair formation increases and becomes length dependent. We shall model the impurities as local obstacles, which pin the dislocation configuration at some points during the process of the kink pair formation. Such modification of the boundary conditions for the calculation of the activation energy brings to the problem solved in Appendix. Its solution is shown in Fig. 6. For long intervals the activation energy $E(\sigma, l)$ is transformed into the result (2.14), for short intervals $E(\sigma, l)$ is given by equation (A.9), which for the present parameters yields

$$E(\sigma, l) \approx \frac{110}{l^5}(1 + 0.3l^2). \tag{5.1}$$

One can see from the Fig. 6 and equation (5.1) that the activation energy strongly increases when $l$ decreases.

Let us consider qualitative features of the activation energy for the limited interval, and return for this purpose to the dimension variables in equation (5.1),

$$E(\sigma, l) \approx \frac{110\kappa^3}{l^5\beta^2}[1 + 0.3\frac{(\beta b\sigma_P\delta)^{1/2}l^2}{\kappa}]. \tag{5.2}$$

The main contribution to $E(\sigma, l)$, $110\kappa^3/(l^5\beta^2)$ is stress independent Therefore, when $\sigma \rightarrow \sigma_P$ the activation energy, modified by the local obstacles, in contrast to the case of long intervals retains finite. Accordingly, the activation volume $\gamma_l = -dE(\sigma,l)/d\sigma$ is determined by the second contribution to $E(\sigma, l)$, equation (5.2),

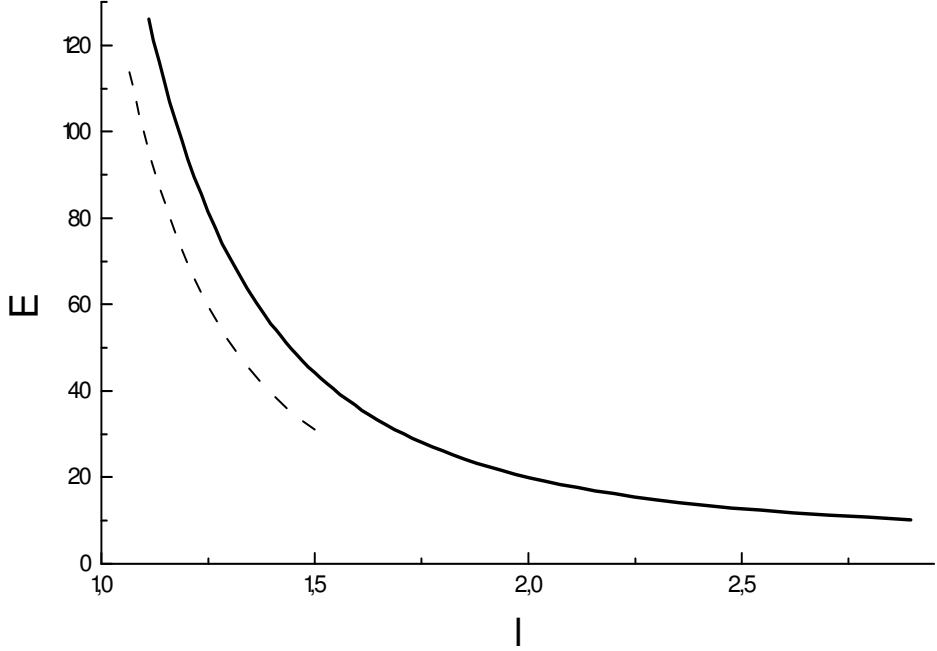

Fig. 6. The energy of the kink pair formation at a restricted interval of length $l$. The dashed line shows the approximation (5.2) for $E(l)$.

$$\gamma_l \approx 16.5 \frac{\kappa^2}{l^3 \beta^{3/2}} \left( \frac{b}{\sigma_P - \sigma} \right)^{1/2}, \qquad (5.3)$$

and contrary to the case of long intervals, it increases as $\sigma \rightarrow \sigma_P$, whereas $\gamma_0$ decreases as $(\sigma_P - \sigma)^{1/4}$, as it follows from equation (2.14). The estimation of the ratio $\gamma_l/\gamma_0$ in the order of magnitude $\gamma_l/\gamma_0 \sim (l_0/l)^3 >> 1$ shows, somewhat unexpectedly, that the activation volume for the kink pair formation at short intervals is larger in comparison with the case of long intervals. This may reveal itself in mechanical tests as a violation of conventional temperature dependencies of the flow stress and the strain rate sensitivity.

The considered model of the influence of the local obstacles is valid until the dislocation segment motion mechanism, operating in the pure material, is qualitative modified. It is supposed, therefore, that the local obstacles are not too strong, or their concentration is not too high in order to change the controlling process (the kink pair formation) to the overcoming local obstacles.

## 6. QUANTUM EFFECTS IN THE THERMOACTIVATION

As is well known, the classical approach to the description of the crystal lattice dynamics is valid at temperatures exceeding the Debye temperature $\Theta \approx \hbar c/(ka)$ of the crystal. Here $\hbar$ and $k$ are the Planck and Boltzman constants, correspondingly, $c$ is the sound velocity, $a$ is the lattice period, which is the shortest wave length of phonons in the system. At $T<\Theta$ quantum effects play a significant role. Similarly, one should expect that quantum properties become apparent in dislocation dynamics at temperatures $T<T_d = \hbar c/(k\lambda)$, where $\lambda$ is a minimal scale for variations of the dislocation configurations. For the case of interest, the most natural choice of $\lambda$ is the kink size $d \approx a(\mu/\sigma_P)^{1/2}$. Thus, a rough estimate for the temperature range of revealing of quantum effects at the dislocation motion in the Peierls-Nabarro relief is $T<T_d \approx \Theta(\sigma_P/\mu)^{1/2}$. This bound for typical metals with the BCC structure is by one or two orders of magnitude lower than the Debye temperature, but modern experimental technique allows to investigate the plasticity of materials at temperatures even below 1K (see, for example, Takeuchi and Maeda 1977 and review (Pustovalov 1989)).

The purpose of the present section is to give a short description of a modification of the thermal activation mechanism of the dislocation motion in the Peierls-Nabarro relief for the range of moderately low temperatures, when quantum effects begin to reveal themselves, but the drastic change of the mechanism itself due to the tunneling does not take place yet. An attention will be also paid to an increasing role of the dislocation viscosity within this intermediate range of temperatures.

The quantum generalization of the calculation of the partition functions $Z_M$ and $Z_0$ is very simple, since for both classic and quantum statistics, they coincide with the partition functions of oscillators with the frequencies $\omega_\alpha = (\mu_\alpha/\rho)^{1/2}$. For negligibly small viscosity ($\eta \to 0$) $Z_\alpha = [2\sinh(\hbar \omega_\alpha/2kT)]^{-1}$ (see, e. g., Kubo 1965). The high temperature limit of it is merely the classic partition function $(Z_\alpha)_{cl} = kT/\hbar \omega_\alpha$. For the considerable viscosity a calculation by the method developed in papers (Caldeira and Leggett 1981, Larkin and Ovchinnikov 1983) yields $Z_\alpha = (Z_\alpha)_{cl} Q_\alpha$, with

$$Q_\alpha = \Gamma(1+n_1^\alpha)\Gamma(1+n_2^\alpha). \tag{6.1}$$

Here $\Gamma(x) = \int_0^\infty t^{x-1} e^{-t} dt$ is the Euler function, and $n_{1,2}^\alpha = (\hbar/2\pi kT)[\eta/(2\rho) \pm (\eta^2/(4\rho^2) - \mu_\alpha/\rho)^{1/2}]$.

Therefore, the employment of the quantum statistics for the description of dislocation fluctuations renormalizes the classic expression for the kink pairs generation rate $\Gamma \to \Gamma_q = \Gamma Q$ ($Q = \prod_\alpha Q_\alpha$) at the low temperature range. Let us estimate the magnitude of this modification.

There is a divergent contribution to $Q$ at $\mu \to \infty$. Since at large $\mu$ the viscosity does not play an essential role (according to the expressions for $n_{1,2}^\alpha$, this takes place at $\mu >> \eta^2/\rho$), one may use the simple expression for $Z_\alpha$ at $\eta = 0$ $Z_\alpha = [2\sinh(\hbar (\mu_\alpha/\rho)^{1/2}/(2kT)]^{-1}$
One has

$$(\ln Q)_{diverg} \approx \int_{\mu_m}^{\mu_M} d\mu [\rho(\mu) - \rho_0(\mu)] \ln[2\sinh(\frac{\hbar}{2kT}\sqrt{\frac{\mu}{\rho}})] \approx \int_{\mu_m}^{\mu_M} d\mu \frac{J}{4\pi\mu^{3/2}} \frac{\hbar}{2kT} \sqrt{\frac{\mu}{\rho}} \approx$$

$$(J\hbar/4\pi kT\rho^{1/2})\ln[\hbar (\kappa/\rho)^{1/2}/(akT)]. \tag{6.2}$$

As the estimation of the upper and low limits of the integration in (6.2) $\mu_M \approx \kappa/a^2$, and $\mu_m \approx \rho(kT/\hbar)^2$ were used, giving the range where the function $\ln(2\sinh[\hbar(\mu/\rho)^{1/2}/(2kT)])$ may be approximated as $\hbar(\mu/\rho)^{1/2}/(2kT)$, and the asymptotic form of $\rho(\mu)$ (2.10) may be employed.

This result enables us to trace the transition to the high temperature case and to obtain quantum corrections to the classical formula for the kink pairs generation rate. These corrections considerably modify the generation rate when $\ln Q$ exceeds unity appreciably. According to (2.16, 6.2), this takes place at temperatures

$$T \leq T_d (3 \cdot 2^{1/2}/\pi\kappa\rho^{1/2})[\beta b(\sigma_P - \sigma)]^{1/4} \ln\{(\kappa^{1/2}/a) [\beta b(\sigma_P - \sigma)]^{-1/4}\}. \tag{6.3}$$

This formula allows verifying the qualitative estimation of the range of revealing of the quantum effects mentioned above. If the influence of the viscosity is not too significant and stresses are not very close to the Peierls stress, one can see from formula (6.3) when substituting the values of parameters that the order of magnitude of the bound temperature $T_d$ is $\Theta(\sigma_P/\mu)^{1/2} \ln(\mu/\sigma_P)^{1/2}$, which differs from the qualitative estimation by the not very essential logarithmic factor slightly increasing $T_d$. However, in the most important for low temperature mechanical tests range of stresses close to the Peierls stress, how it can be seen from formula (6.3), $T_d$ slightly decreases in comparison with the above mentioned estimation and it is approximately $T_d \approx \Theta(\sigma_P/\mu)^{1/2} ((\sigma_P - \sigma)/\sigma_P)^{1/4} \ln[(\sigma_P/\mu)^{1/2} ((\sigma_P/(\sigma_P - \sigma_P))^{1/4}]$. This follows from the fact that the characteristic size of the dislocation configuration $y_0(x)$ exceeds the kink size just by the factor $((\sigma_P/(\sigma_P - \sigma_P))^{1/4}$. Due to the small value of the exponent, (1/4), this additional factor does not cause a significant decrease in $T_d$ over a wide range of stresses.

The next contribution of interest in $\ln Q$ is connected with the viscosity. Let us remind that in the classical limit there is only the pre-exponential viscosity dependence of $\Gamma$ (trough $\Omega$, see (3.1)). Therefore, the arising due to $Q$ an exponential dependence on $\eta$ provides an enhancement of the viscosity dependence of a purely quantum mechanical origin. This effect is the most pronounced at $T \to 0$, and, for ordinary materials, within the stress range close to the Peierls stress. For this case the complete expression for $Q$ was found in (Petukhov 1986):

$$Q = Q_{cont} \Gamma(1 + z\bar{\eta})\Gamma(1 + z(\bar{\eta}/2 + \sqrt{\bar{\eta}^2/4 - 3/2}))\Gamma(1 + z(\bar{\eta}/2 - \sqrt{\bar{\eta}^2/4 - 3/2})) \times$$

$$\frac{\Gamma(1 + z(\bar{\eta}/2 + \sqrt{\bar{\eta}^2/4 + 5/2}))\Gamma(1 + z(\bar{\eta}/2 - \sqrt{\bar{\eta}^2/4 + 5/2}))}{[\Gamma(1 + z(\bar{\eta}/2 + \sqrt{\bar{\eta}^2/4 - 2}))\Gamma(1 + z(\bar{\eta}/2 - \sqrt{\bar{\eta}^2/4 - 2}))]^3}. \tag{6.4}$$

Here $Q_{cont}$ is the contribution from the continuous spectrum,

$$Q_{cont} = \prod_{n=1}^{N_m} (1 + \frac{\sqrt{2}z}{\sqrt{n^2 + n\bar{\eta}z + 2z^2}})^2 (1 + \frac{z/\sqrt{2}}{\sqrt{n^2 + n\bar{\eta}z + 2z^2}})^2 (1 + \frac{3z/\sqrt{2}}{\sqrt{n^2 + n\bar{\eta}z + 2z^2}})^2, \tag{6.5}$$

$\bar{\eta}$ is the dimensionless viscosity $\bar{\eta} = \eta/[\rho^{1/2}(\beta b \delta)^{1/4}]$, $z = \hbar (\beta b \delta)^{1/4}/(2\pi kT\rho^{1/2})$

To illustrate the viscosity dependence of $\Gamma$ at small $\eta$, we extract from $\ln Q_{cont}$ the term $\bar{\eta} \varphi(z)$ linear in $\eta$. By expansion of (6.5) on $\eta$ we get

$$\varphi(z) = (z^2/2) \times \sum_{n=1}^{\infty} \left( \frac{\sqrt{2}n}{(\sqrt{n^2+2z^2}+\sqrt{2}z)(n^2+2z^2)} + \frac{n/\sqrt{2}}{(\sqrt{n^2+2z^2}+z/\sqrt{2})(n^2+2z^2)} + \right.$$
$$\left. \frac{3n/\sqrt{2}}{(\sqrt{n^2+2z^2}+3z/\sqrt{2})(n^2+2z^2)} \right). \tag{6.6}$$

Simplified analytical approximation of this function may be given as

$$\varphi(z) = -z^2(19z+5.5)(9.76z^2+1.8z+1). \tag{6.7}$$

One can see from equation (6.7) that in the classical limit $T \to \infty$ ($z \to 0$), $\varphi(z) \to 0$, and the exponential viscosity dependence of $\Gamma$ disappears. However, for $z \sim 1$ the coefficient at $\eta$ in $\ln Q_{cont}$ is large, and, therefore, the viscosity dependence is strong.

In the case of the large viscosity the estimation of $\mu_m$ in equation (6.2) should be changed into $\mu_m \approx \eta^2/\rho$, and the expression for $\ln Q_{diverg}$ takes the form

$$\ln Q_{diverg} \approx (\hbar J/4\pi kT\rho^{1/2})\ln((\kappa\rho)^{1/2}/\eta a). \tag{6.8}$$

If the viscosity is temperature independent (for example, determined by the electronic component (see, e. g., Kaganov, Kravchenko and Natsik 1974), then, owing to the linear dependence of $\ln Q$ on the inverse temperature, its contribution may be considered as the viscosity-dependent quantum renormalization of the classical kink pair activation energy $E = E_0 + \delta E$, where $\delta E$ is

$$\delta E = -(\hbar J/4\pi\rho^{1/2})\ln((\kappa\rho)^{1/2}/\eta a). \tag{6.9}$$

A more strong viscosity dependence of the dislocation mobility, than that predicted by the classical Kramers formula for the rate of the thermally activated processes (3.1), was observed in experiments with the destruction of the superconductive state by the magnetic field in Nb (Karpov, Leiko and Nadgorny 1980) and Sn (Natsik *et al*. 1996).

The theory presented describes the quantum effects on the thermally activated motion of dislocations. At sufficiently low temperatures the mechanism of thermally activated overcoming barriers should change into the quantum tunneling. As an indication to such modification of the mechanism in the present theory serves the singular behavior of the partition function, which corresponds to the pass coordinate (see (6.1)),

$$Z_0 \propto \Gamma\{1+(\hbar/(2\pi kT))[(\eta/2\rho)-((\eta/2\rho)^2+|\mu_0|/\rho)^{1/2}]\} \to \infty \text{ as}$$
$$T \to T_c = (\hbar/2\pi k)[((\eta/2\rho)^2+|\mu_0|/\rho)^{1/2}-(\eta/2\rho)]. \tag{6.10}$$

## 7. TUNNEL CHANNEL OF KINK PAIR GENERATION

At the present time the temperature range for the mechanical tests of materials becomes considerably wider. A great interest of investigators is paid now to the studying of the plastic deformation of materials at extremely low temperatures, when some new features of the fundamental mechanisms of the dislocation dynamics reveal themselves in a number of peculiarities (see, e. g., Pustovalov 1989). One of this peculiarities is a change of nature of the temperature dependence of the plastic flow.

Observations of a weakening of the temperature dependence of parameters of the plasticity (the athermal behavior) have a long history. The evolution of the experimental investigations of the low temperature plasticity of materials has been described in the review paper (Pustovalov 1989). Among recent papers one should note (Brunner and Diehl 1992, Natsik et al. 1996). Due to the substantial progress in expanding of the experimental investigations toward the zero temperature a new area with the drastically different behavior of BCC metals, alkali halides, and some other materials was discovered. In this area the strong temperature dependence of the plastic flow of the Arrhenius type weakens and changes to an athermic one. The appearance of quantum mechanical tunnelling is considered beginning since the Mott paper (Mott 1956) as the most fundamental reason for the athermal behavior. The probability of quantum fluctuations does not depend on the temperature. Accordingly, to account for the quantum mechanical effects the drastic decrease of the thermally activated dislocation mobility at $T\rightarrow 0$ must be changed by the athermic regime. This transition provides a rare opportunity to observe the quantum mechanical effects on a macroscopic level by mechanical experiments.

For sufficiently light materials, for example, LiH, in which the average mass of atoms is close to the mass of helium atoms, one may expect a considerable amplitude of zero-point vibrations. In this case the quantum mechanical nature of the athermal behavior of the plasticity seems to be obvious. For the identification of mechanisms of anomalies in more heavy materials a detailed quantitative description of quantum mechanical effects in dislocation motion is desirable.

Sometimes for the generalized semiqualitative description of the dislocation mobility in the transitional to the quantum regime range one uses a concept of the "effective temperature". The effective temperature is introduced in analogy with the substitution, which has to be done in the expression for the probability $W$ of one-dimensional oscillators deviations $q$

$$W=\exp\{-m\omega^2 q^2/2kT\} \qquad (7.1)$$

in order to make this formula valid in the quantum case also (see, e. g., Landau and Lifshitz 1958 b)

$$T\rightarrow T^*=(\hbar\omega/2k)\coth(\hbar\omega/2kT). \qquad (7.2)$$

Using the effective temperature $T^*$ instead of $T$ in the Arrhenius law one can to describe correctly the high temperature classical limit $T\rightarrow\infty$, when $T^*$ coincides with $T$, and yields also the weakening of the temperature dependence in the low temperature limit $T\rightarrow 0$, when $T^*\approx\hbar\omega/2k$. For the characteristic frequency $\omega$ the estimation $\omega=(2\pi\sigma_P a/m)^{1/2}$ was used, which allows to describe qualitatively the observed deviations from the conventional Arrhenius law for typical BCC metals, where $\omega\approx 10\ K$ (Takeuchi and Hashimoto 1984). However, the real mode of the dislocation motion over barriers may differ significantly from the one-dimensional harmonic oscillations. In the following a more detailed consideration taking into account the multidimensional nature of the dislocation motion within the framework of the string model will be given.

For the most of materials the quantum effects reveal themselves in the range of low temperatures and stresses close to the Peierls stress. This case will be discussed more thoroughly. However, as an Introduction, some results for other stress range will be first presented, which are of interest mainly in relation to specific quantum crystals. For example, for solid helium or hydrogen, in which the dimensionless parameter $\hbar/(\kappa\rho)^{1/2}a$, proportional to the square of the ratio of the zero-point vibrations amplitude to the lattice period and characterizing the quantum properties of materials, is not too small.

First qualitative consideration of the dislocation tunnelling trough the Peierls-Nabarro barriers was given by Weertman (Weertman 1968). Weertman modeled a dislocation as a "particle", and used the well known quasiclassical formula for the tunnel transition trough a one-dimensional potential barrier $U(x)$ (see, e. g., (Landau and Lifshitz 1958b))

$$\Gamma=\exp\{-S_0/\hbar\}, \tag{7.3}$$

with the action

$$S_0=2\int\sqrt{2mU(x)}dx. \tag{7.4}$$

Weertman used, on the intuitive base, the next parameters of the "particle": $m$ equals to the mass of an atom of the material, the height and the width of the barrier equal to $2E_k$ and $b$, correspondingly. He came to the result

$$\Gamma\approx\exp\{-(4(mE_k)^{1/2}b/\hbar\}. \tag{7.5}$$

A similar approach was developed by Mil'man and Trefilov in (Mil'man and Trefilov 1966), where, from the comparison of rates of the quantum tunnelling and the classical thermally activated overcoming the Peierls-Nabarro barrier the temperature $T_c$ of the transition to the athermic regime was estimated as $T_c \lesssim (\hbar/\pi ak)(E_k/m)^{1/2}$. More consistent theory of the dislocation tunnelling for the overcoming local obstacles was developed in (Natsik 1983, Natsik and Rotschupkin 1980, Vardanian and Ossipian 1988, etc.), and for the overcoming the crystal relief in (Petukhov and Pokrovskii 1973, Suzuki 1982, Nakaya and Hida 1986, Petukhov, Koizumi and Suzuki 1998, etc.).

At present, it is generally recognized that the dislocation motion mode over the Peierls-Nabarro barrier at low stress is the kink pair formation and expansion. This makes obvious, which parameters should be substituted into the tunnel transition formula (7.3). The tunnelling "particle" is an expanding kink pair, and the potential barrier is given by

$$U(x)=2E_k-\sigma abx, \tag{7.6}$$

where $x$ is the kink pair size.

The corresponding mass is one for the relative kink motion $m_k/2$, where $m_k=\rho E_k/\kappa$ is the mass of a kink, $\rho$ is the mass density per a unit length of the dislocation.

Calculating the integral in (7.4), one obtains

$$\Gamma=\exp\{-\sigma_q/\sigma\} \tag{7.7}$$

with $\sigma_q=(8E_k^2/3\hbar ab)(2\rho/\kappa)^{1/2}$.

The most important difference from the Weertman result (7.5) is a strong stress dependence of the transition rate. Later the same force dependence of the tunnel nucleation of solitons in a similar problem of the charge density wave motion was obtained by Maki (Maki 1978). In this paper the "relativistic" two-dimensional symmetry of the problem was taken into account. Maki exploited an analogy between 1-dimensional quantum mechanical problems and 2-dimensional statistical ones, in which the tunnel transition of a string corresponds to the formation of a nucleus of a stable phase in two-dimensional metastable media.

In such approach the action plays the role of the "free energy", which consists of the "boundary" term, determined by the kink energy $E_k$, and the "surface" term, determined by an external force. Due to the similarity of the boundary conditions at $x$ and $t$ axes in the case $T=0$, this "free energy" depends only on one mixed coordinate $r$, the radius of the nucleus in the two-dimensional space-time. In our notation, taking into account the scaling factor between the time and the space coordinates $t=(\rho/\kappa)^{1/2}x$, we have

$$S=(\rho/\kappa)^{1/2}(E_k 2\pi r - \sigma ab \pi r^2) \tag{7.8}$$

The calculation of the extreme of $S$ over $r$ yields a more exact numerical factor for $\sigma_q$ in (7.7)

$$\sigma_q=(\pi E_k^2/\hbar\, ab)(\rho/\kappa)^{1/2}, \tag{7.9}$$

The quantum-mechanical theory of the fluctuational breakaway of kinks from pinning centers was developed in (Hovakimian, Kojima, and Okada 1995). The dislocation motion in quantum crystals under the action of an oscillating force was considered in (Markelov 1985). The effect of dissipation on the dislocation kinks tunnelling was discussed in (Hikata and Elbaum 1985).

## 8. COMBINED ACTIVATED-TUNNEL MECHANISM OF DECAY OF METASTABLE STATES

Now let us describe shortly, as an introduction to the general case, the one-dimensional picture of the quantum mechanical barrier overcoming at finite temperatures. At $T>0$ the tunnelling proceeds with a preliminary activation by some amount of energy $E$. In such the case, one should use equation (7.3) with the barrier height reduced by $E$

$$\Gamma(E)=\exp\{-S_0(E)/\hbar\}, \tag{8.1}$$

where

$$S_0(E)=2\int\sqrt{2m[U(x)-E]}dx.$$

The total rate of the transition is the sum of contributions over the different energies taken with the Boltzman factors $\exp\{-E/kT\}$, accounting for the distribution of states of the "particle" over energy in front of the barrier

$$\Gamma(T)\propto\int\exp\{-\frac{E}{kT}-\frac{S_0(E)}{\hbar}\}dE. \tag{8.2}$$

The main contribution to the integral (8.2) comes from the energy corresponding to the maximum of the exponent, which is determined by the extreme condition

$$-\frac{dS_0(E)}{dE} = \frac{\hbar}{kT}. \tag{8.3}$$

The final expression for the transition rate at finite temperatures (8.4) does not bear any specific features connected with the one-dimensionality, and it is valid in more general multidimensional case as well. This fact, which we leave without of the rigorous proof, will be used in the following.

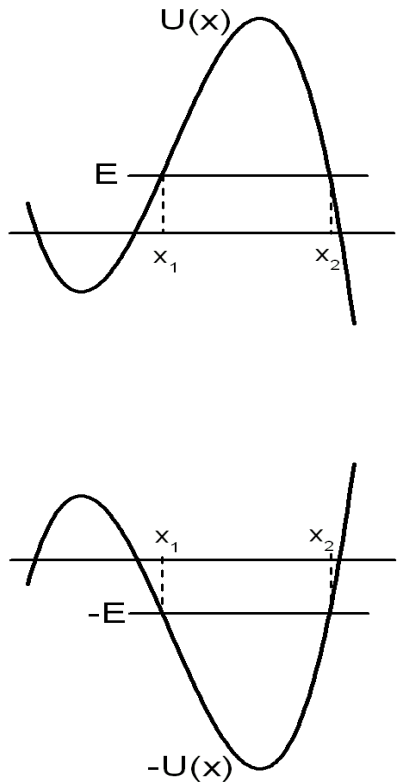

Fig. 7. Illustration of the quasiclassical motion in the inverted potential. $E$ is the most probable energy of preactivation.

It is known from the ordinary classical mechanics (see, e. g., Landau and Lifshitz 1958a) that the quantity $-dS_0(E)/dE$ is the period $t(E)$ of the motion over a trajectory (in inverted potential $U(x) \rightarrow -U(x)$) with the given energy (see Fig. 7). Therefore, the quasiclassical transitional trajectory corresponds to the motion with the period of time $\hbar/kT$. The total action $S=S_0(E)+Et$ can be represented as the integral from the so-called 'Euclidean' Lagrangian $L_E$ (i.e. with the potential inverted) of the system $S=\int L_E dt$, and we arrive to the answer for the transition rate

$$\Gamma(T) = \exp\left\{ -\frac{1}{\hbar} \int_{-\hbar/2kT}^{\hbar/2kT} L_E dt \right\} \tag{8.4}$$

Equation (8.3) has a solution in the range $0 \leq E \leq U$, where $U$ is the height of the barrier. Near the top of the barrier a quadratic approximation of any reasonable potential $U(x)$ is valid $U(x) \approx U - \mu_1 x^2/2$. Thus, for $E \to U$ we have

$$S_0(E) \approx 2 \int [2m(U-E-\mu_1 x^2/2)]^{1/2} dx = 2\pi(U-E)(m/\mu_1)^{1/2} \tag{8.5}$$

For typical potentials the derivative of this expression over $E$ determines a possible minimum of the period of the motion along the transitional trajectory $-dS_0/dE = 2\pi(m/\mu_1)^{1/2}$, which, according to equation (8.3), yields an upper boundary

$$T_c = \frac{\hbar}{2\pi kT} \left( \frac{\mu_1}{m} \right)^{1/2} \tag{8.6}$$

of the temperature range in which a solution of equation (8.3) exists. At $T \to T_c$ we have $E \to U$, and $\Gamma \to \exp(-U/kT)$, that is the classical rate of the thermally activated transition (the Arrhenius factor).

## 9. QUANTUM DIFFUSION OF KINKS IN THE PEIERLS-NABARRO RELIEF OF SECOND KIND

The one-dimensional consideration of the preceding section may be directly applied for the description of the kink migration in the secondary relief. Our aim now is to determine the temperature dependence of the coefficient $D_k$ of kink diffusion along the dislocation line. At high migration barriers, the kinks formed are localized mainly in the vicinity of the secondary relief minima, and the displacement of a kink from one minimum to another occurs at high temperature due to thermal fluctuations. Then, according to the Arrhenius law, $D_k = D_0 \exp(-E_m/kT)$, where $E_m$ is the height of kink migration barrier, $D_0$ is the preexponential factor inessential for further consideration estimated in (Lothe and Hirth 1959) as $D_0 \sim \nu_D a_2^2$ ($\nu_D$ is the Debye frequency, $a_2$ is the period of the second kind Peierls-Nabarro relief). We shall model a kink by a particle of mass $m_k$ moving in the inclined potential $U(x) = U_2(x) - Fx$ (where $U_2(x)$ is the secondary relief, $F = \sigma ab$ is the driving force). We assume the kink mass to be constant irrespectively of the coordinate or motion velocity. We perform the calculations for a specific potential of form $U_2(x) = (E_m/2)[1 - \cos(2\pi x/a_2)]$.

For the boundary between the classical and quantum temperature ranges we have from (8.6) at $Fa_2 << E_m$

$$T_{c2} = \frac{\hbar}{ka_2} \sqrt{\frac{E_m}{2m_k}} \left[ 1 - \left( \frac{Fa_2}{\pi E_m} \right) \right]^{1/4} \approx \frac{\hbar}{ka_2} \sqrt{\frac{E_m}{2m_k}} . \tag{9.1}$$

Now determine the temperature dependence of kink mobility at $T<T_{c2}$. Neglecting the effect of the external force $F$ on the shape of the migration barrier, we can write the action $S(E)$, determining the probability of the tunnel transition over period as an integral

$$S(E) = 2\int_{x_1}^{x_2} \sqrt{2m_k[U_2(x)-E]}dx. \qquad (9.2)$$

Here $x_1$ and $x_2$ are turning points of the transitional trajectory (see Fig. 7). Using the explicit form of the potential $U_2(x)$, we can express $S(E)$ in terms of the tabulated functions

$$S(E)=S(0)[E(1-E/E_m)-(E/E_m)K(1-E/E_m)]. \qquad (9.3)$$

Here $S(0)=(4a_2/\pi)(2m_k E_m)^{1/2}$ is the value of $S(E)$ in tunnelling without preliminary activation (with zero energy),

$$K(m) = \int_0^{\pi/2} \frac{d\theta}{\sqrt{1-m\sin^2\theta}}$$

and

$$E(m) = \int_0^{\pi/2} \sqrt{1-m\sin^2\theta}\,d\theta$$

are the complete elliptic first- and second-kind integrals, respectively (Abramovitz and Stegun 1964).

In a similar way, equation (8.3) for determination the optimum preactivation energy can be written as

$$K(1-E/E_m)=2\hbar E_m/kTS(0). \qquad (9.4)$$

Determining $E$ as a function of temperature $E(T)$ from this equation and calculating

$$A(T)=E(T)/kT+S(E(T))/\hbar, \qquad (9.5)$$

we arrive at the temperature dependence of the diffusion coefficient

$$D_k \sim \exp[-A(T)]. \qquad (9.6)$$

Formulae (9.3)-(9.5) can be considered as the parametric representation of the temperature dependence of $A$ ($E$ is the parameter). The behavior of the diffusion coefficient (9.6) is shown in Fig. 8. The straight line corresponds to the Arrhenius dependence in the high-temperature range. At $T<T_{c2}$ there are deviations from the Arrhenius dependence caused by the contribution of the tunnel mechanism of the barrier overcoming. At $T<<T_{c2}$ the coefficient of kink diffusion has a constant athermic value $D_k \sim \exp(-S(0)/\hbar)$.

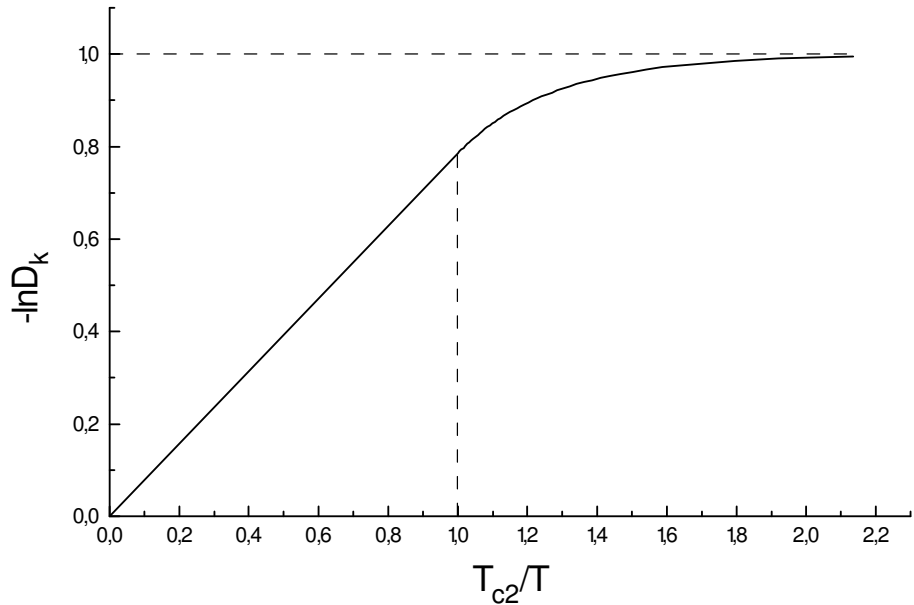

Fig. 8. Logarithm of the kink diffusivity (in units of $S(0)/\hbar$) as a function of reciprocal temperature $T_{c2}/T$.

As a rule, the crystals with a high secondary relief are brittle in the low temperature range. In such materials, the range of manifestation of the tunnel mechanism of kink formation can hardly be attained. At the same time, the temperature range of quantum-effect manifestation during kink migration can extend up to the Debye temperature of crystals and is accessible for the studying the mechanical properties of crystals. Especially useful may be technique of low-temperature testing of materials in conditions of the hydrostatic compression preventing from the cracks formation.

Now we have the sufficient insight in the quantum theory of the decay of metastable states in order to turn to the main problem of interest, the tunnelling of an extended dislocation through a Peierls-Nabarro barrier.

## 10. TUNNELLING IN MULTIDIMENSIONAL SPACE

For a calculation of the transition time $t_{tr} \sim 1/\Gamma$ in the quantum regime we shall use equation (8.4), bravely generalizing it in two ways. The first generalization is to use it for the manydimensional system, the dislocation, considered as a flexible elastic string. The second generalization is the accounting for the dissipative quantum dynamics. However, we shall first consider in detail more simple case of negligible small dissipation $\eta \to 0$.

The corresponding Lagrangian contains, besides the potential energy (2.1), also the kinetic energy

$$L_E = \int_{-\infty}^{\infty} dx \left[ \frac{\rho}{2}\left(\frac{\partial y}{\partial t}\right)^2 + \frac{k}{2}\left(\frac{\partial y}{\partial x}\right)^2 + U(y) \right], \tag{10.1}$$

where $y(x,t)$ describes the position of the dislocation in a slip plane at point $x$ and time $t$. Compared with the one-dimensional case, the novelty of equation (10.1) is that the coordinate $y$ is not a function of time alone but it depends also on $x$, the coordinate along the dislocation line. The problem to be solved is the minimization of the functional of action in (8.4) with respect to a trajectory transferring the dislocation through the barrier.

We shall consider now the range of stresses in the vicinity of the Peierls stress, when the barrier is considerably reduced by the external loading. Rescaling the variables $x$, $y$, $t$ to new dimensionless values: $y \to = y(\beta/\sigma_P b \delta)^{1/2}$, $x \to x(\beta b \sigma_P \delta)^{1/4}/\kappa^{1/2}$, $t \to t(\beta b \sigma_P \delta)^{1/4}/\rho^{1/2}$ ($\delta = 1 - \sigma/\sigma_P$), we may rewrite the dimensionless action

$$\bar{S} = (\beta/\rho^{1/2} b \sigma_P \delta) S = \int_{-\pi z}^{\pi z} dt \int_{-\infty}^{\infty} dx \left[\frac{1}{2}\left(\frac{\partial y}{\partial t}\right)^2 + \frac{1}{2}\left(\frac{\partial y}{\partial x}\right)^2 + y^2 - \frac{1}{3} y^3 \right] \tag{10.2}$$

as a function of only one parameters, the dimensionless (inverse) temperature $z = \hbar(\beta b \sigma_P \delta)^{1/4}/(2\pi k T)$. This combined variable has a simple meaning: the natural scale for the temperature is the temperature boundary between the quantum and the classical regimes,

$$T_c = \sqrt{\frac{5}{8}} \frac{\hbar}{\pi k \rho^{1/2}} (\beta b \sigma_P \delta)^{1/4}, \tag{10.3}$$

and $z = (2/5)^{1/2} T_c/T$. Note that expression for $T_c$ (10.3) agrees with the general formula for $T_c$ (6.10) at $\eta = 0$, and with the one-dimensional estimation (8.6) at the replacement $\mu_1/m \to |\mu_0|/\rho = 2.5(\beta b \sigma_P \delta)^{1/4}/\rho$.

This scaling transformation allows to obtain some results immediately, without of calculations (Petukhov and Pokrovskii 1973). First, in the case of zero temperature, $T \to 0$, when $z \to \infty$, $\bar{S}$ becomes simply a numerical factor, $S_0$, and the dependence of the action $S$ on all parameters is given directly by the scale factor

$$S = S_0 (\kappa \rho)^{1/2} b (\sigma_P - \sigma)/\beta. \tag{10.4}$$

Variational (Petukhov 1985a) and numerical calculations (Nakaya and Hida 1986) yield for $S_0$ the estimation $S_0 \approx 31$.

Result of numerical calculations of the action for the harmonic Peierls- Nabarro potential (1.1) at $T=0$ in the total stress range $0 < \sigma < \sigma_P$ (Nakaya and Hida 1986) is shown in Fig. 9. One can approximate the obtained stress dependence by an analytic expression interpolating between

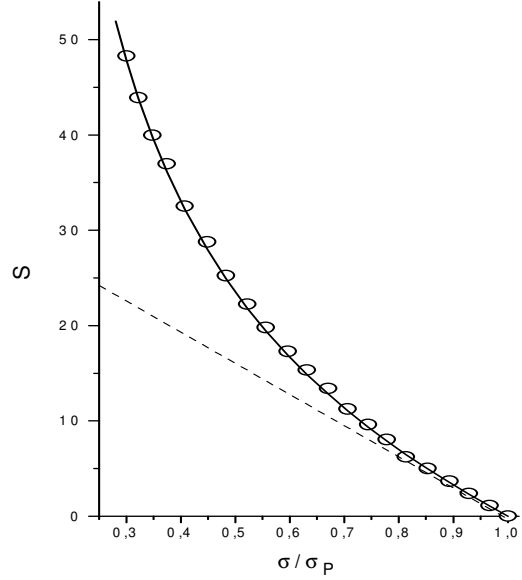

Fig. 9. The stress dependence of action for a wide range. Circles show results of numerical calculations for the harmonic potential (Nakaya and Hida 1986). The dashed line represents the high-stress behavior (10.4).

low-stress (7.7) and high-stress (10.4) behavior

$$S(\sigma)/\hbar = q[16\sigma_P/\sigma - 1 - 15\sigma/\sigma_P], \tag{10.5}$$

where $q = (\kappa\rho)^{1/2} a^2 / 2\pi^2 \hbar$.

It is also possible to derive some general relation between experimentally measured quantities, such as the temperature and strain rate sensitivities of the flow stress $\sigma(T)$, replacing the well known relation of the thermally activated analysis (4.2). In the quantum regime at $\sigma \to \sigma_P$ this classical relation is replaced by the following one, which is a consequence of the dependence of $\bar{S}$ only on one combined variable $(\sigma_P - \sigma)^{1/4}/T$ (Petukhov 1985a)

$$\ln(\dot{\varepsilon}_0/\dot{\varepsilon}) = \frac{\sigma_P - \sigma + \dfrac{T}{4}\partial\sigma/\partial T}{\partial\sigma/\partial\ln\dot{\varepsilon}}. \tag{10.6}$$

In particular, as $T \to 0$ it follows from (10.6) that $\delta\sigma_0 = \sigma_P - \sigma(0) = (\ln\dot{\varepsilon}_0/\dot{\varepsilon})(\partial\sigma/\partial\ln\dot{\varepsilon})(T=0)$.

To describe the temperature dependence of the dislocation mobility completely, one has to find the optimal action $\bar{S}$ as a function of $T$ in the total quantum interval $0 < T < T_c$. The variational method proves to be effective for the solving of this problem (Petukhov 1985, Petukhov, Koizumi and Suzuki 1998). At finite $T$ the time interval of the motion along the transitional trajectory is limited, and there is no more the "relativistic" symmetry between time and space coordinates. We shall use for the determination of the action a multiplicative trial function

$$y(x,t) = \lambda(t) q(x), \tag{10.7}$$

where both functions, $\lambda(t)$ and $q(x)$ are variational parameters, and they are selected in an optimal way. The factorization of equation (10.7) allows to separate effectively the time and space problem. In general it is possible to find an optimal shape $q(x)$ for any function $\lambda(t)$, and, therefore to remove the space variable from the problem. For this purpose we substitute equation (10.7) into equation (10.1) and obtain for the space dependent function $q(x)$ the problem of type considered above, equations (2.1), (2.12), but with different parameters, depending on $\lambda(t)$ and undetermined yet. Nevertheless, its solution has the same functional form (2.13) (see also Appendix, equation (A.7))

$$q(x) = \frac{1}{\cosh^2(Bx)}, \tag{10.8}$$

where the parameter $B$, which governs the space scale of the tunnelling nucleus, will be determined later in result of the over-all optimization. The substitution of this $q(x)$ into equation (10.1) allows reducing the problem to the effectively one-dimensional one

$$\bar{S} = \int_{-\pi z}^{\pi z} [\frac{B_1}{2}\dot{\lambda}^2 + \frac{B_2}{2}\lambda^2 - \frac{B_3}{3}\lambda^3]dt, \tag{10.9}$$

where the coefficients $B_1=4/(3B)$, $B_2=16B/15+8/(3B)$, $B_3=16/(15B)$ depend on the shape of the tunnelling nucleus through the uncertain yet parameter $B$.

The solution of the problem for the functional (10.9) is similar to that given in Appendix, with the interval length having now the meaning of the time period $2\pi z$. Calculating $\bar{S}$ as a function of $B$ and finding its minimum over all possible values of $B$, we arrive to the final answer. The result of the numerical calculation is presented in Fig. 10. In this figure also an approximation of the numerical data by the analytical function

$$\bar{S}(z) = P_0 - P_1 \exp(-P_2 z) \tag{10.10}$$

with $P_0=31.25$, $P_1=4011$, and $P_2=10.83$ is shown.

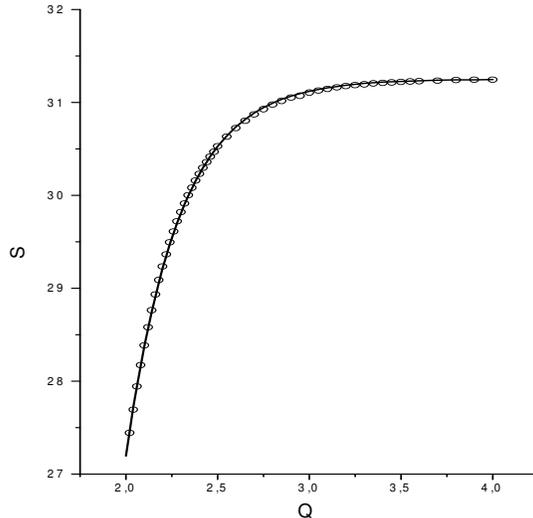

Fig. 10. The temperature dependence of the action. Circles show results of numerical calculations, solid line represents the fitting dependence (10.10).

In practice, these tedious variational calculations are necessary only to establish the functional form of the temperature dependence of the action. If it is taken on trust, then one may obtain even a slightly better approximation using for the determination of the constant in (10.10) the exact value of $\bar{S}$ at $T=T_c$, $S_c=96\pi/5^{3/2}\approx 27$, which is known from the matching with the Arrhenius law, and the better estimation of $\bar{S}$ at $T=0$, $S_0\approx 31$, arriving to the result

$$\bar{S} = S_0 - (S_0 - S_c)\exp\{\frac{S_c}{S_0 - S_c}(1 - \frac{T_c}{T})\}. \tag{10.11}$$

## 11. DISSIPATIVE QUANTUM DYNAMICS OF DISLOCATIONS

Since the energy dissipation plays an important role in dislocation dynamics (Kaganov, Kravchenko and Natsik 1973, Alshits 1992) it is necessary to take it into account also in consideration of the processes of barriers overcoming. Now we shall investigate the effect of the viscosity on the dislocation tunnelling through the Peierls-Nabarro barriers.

Following the method of incorporating dissipation into the quantum mechanics developed for $T=0$ by Caldeira and Leggett (Caldeira and Leggett 1981) and generalized for finite temperatures by Larkin and Ovchinnikov (Larkin and Ovchinnikov, 1983a, 1983b, 1984) and Waxman and Leggett (1985), we include into the 'Euclidean' Lagrangian $L_E$ the viscous term

$$\frac{\eta}{4\pi}\int_{-\infty}^{\infty}dx\int_{-\infty}^{\infty}dt'\left(\frac{y(x,t)-y(x,t')}{t-t'}\right)^2 \tag{11.1}$$

In result, our dimensionless action $\bar{S}$ becomes a function of two variables: $z$ and the dimensionless viscosity $\bar{\eta} = \eta/(\rho^{1/2}(\beta b\sigma_P\delta)^{1/4})$.

There are numerous calculations of action $S$ for different ranges of parameters: for large viscosity (Petukhov 1985), for small and large viscosity at $T=0$ (Nakaya and Hida 1986), for low stresses (Ivlev and Mel'nikov 1987), etc., giving separate representation of results, partly analytical, partly as plots. This makes applications of these results to the description of experimental data inconvenient, and there is a necessity in simplified united formulas. One variant of interpolation between different results was suggested in (Natsik *et al.* 1996). We would like to present another variant of the united description, which complements results of new calculations (Petukhov, Koizumi and Suzuki 1998). It is based on the generalization of the temperature dependence given by equation (10.11)

$$\bar{S}(T) \approx S_0(\bar{\eta}) - [S_0(\bar{\eta}) - S_c(\bar{\eta})]\exp\left\{(1 - T_c(\bar{\eta})/T)\frac{S_c(\bar{\eta})}{S_0(\bar{\eta}) - S_c(\bar{\eta})}\right\}. \tag{11.2}$$

Here $S_0(\bar{\eta})$ is the action at $T=0$, for which, in its turn, the approximation

$$S_0(\bar{\eta}) \approx 31[(1+0.067\bar{\eta}^2)^{1/2} + 0.55\bar{\eta}] \tag{11.3}$$

is suggested. $S_c(\bar{\eta})$ is the action at the boundary with the thermally activated regime $T=T_c(\bar{\eta})$

$$T_c(\bar{\eta}) = T_P \delta^{1/4}[(5/2 + \bar{\eta}^2/4)^{1/2} - \bar{\eta}/2], \qquad (11.4)$$

$$S_c(\bar{\eta}) = 48 \cdot 2^{1/2} \pi/[5((5/2 + \bar{\eta}^2/4)^{1/2} - \bar{\eta}/2)]. \qquad (11.5)$$

The expression (11.2) approximates well the temperature dependence of the action for small and moderate values of the viscosity. Its characteristic feature is the smooth matching with the classical Arrhenius exponent. The case of the large viscosity $\bar{\eta} \gg 1$ can be separately approximated by equation, resulting in this case from the general formulae of the paper (Petukhov, Koizumi and Suzuki 1998)

$$\bar{S} = (8\pi/3)(5/3)^{3/2} \bar{\eta} [1 - 3\theta/2 + (1-\theta)^{3/2}]^{1/2}, \qquad (11.6)$$

where $\theta = (\pi k T \eta/2\hbar)^2/(\beta b \sigma_P \delta)$.

## 12. COMPARISON WITH EXPERIMENTAL DATA

The detailed review of the experimental investigations of the athermic anomalies of the flow stress in materials with the Peierls-Nabarro mechanism of the dislocation motion was given in (Pustovalov 1989), and we restrict ourselves only by a few typical examples.

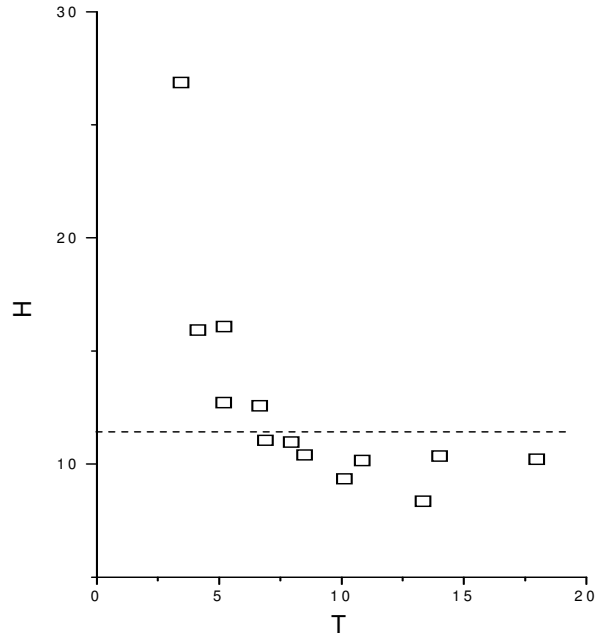

Fig. 11. The check of the Arrhenius law for KCl. Squares represents the right hand of equation (4.4) calculated with the using data (Suzuki and Koizumi 1985).

Taking into account that the kink formation process rate $\Gamma$ is given by equation (8.4) in the quantum regime and by the Arrhenius factor $\exp(-E_0(\sigma)/kT)$ in the classical regime, we have

$$\ln(\frac{\dot{\varepsilon}_0}{\dot{\varepsilon}}) = \begin{cases} S(\sigma, T)/\hbar, & T < T_c \\ E_0(\sigma)/kT, & T > T_c \end{cases}. \qquad (12.1)$$

This relationship predicts a violation of the Arrhenius law below a certain temperature. An example of a check of this prediction with the use of equation (4.4) for KCl is shown in Fig. 11.

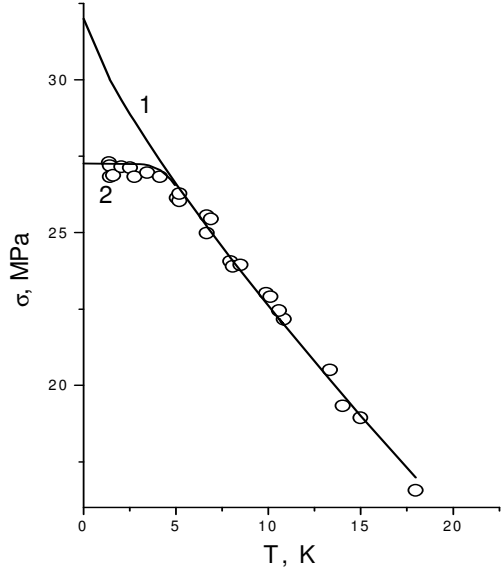

Fig. 12. The temperature dependence of the flow stress for KCl. Circles represent data (Suzuki and Koizumi 1985). The line 1 shows the classical dependence (4.3), the line 2 represents the result of calculations taking into account the quantum tunneling of dislocations.

The experimental data (Suzuki and Koizumi 1985) for the right hand of equation (4.4) are plotted as a function of temperature. One may see that for $T>7\ K$ the plotted quantity is, within the limits of experimental error, temperature independent, and the criterion (4.4) of operation of the thermally activated kinetics fulfils. However, for $T<7K$ there is a large systematic increase of this quantity with the decrease of temperature. This temperature anomaly clearly evidences about a violation of the thermally activated Arrhenius type kinetics. Let us try to describe the experimental data at $T\rightarrow 0$ using the quantum kinetics.

At $T \leq T_c$ the solution of equation (12.1) may be represented as

$$T = T_c \left[ 1 + \frac{S_0 - S_c}{S_c} \ln \left( \frac{S_0 - S_c}{S_0 - \frac{\hbar\beta \ln(\dot{\varepsilon}_o / \dot{\varepsilon})}{(\kappa\rho)^{1/2} b(\sigma_P - \sigma)}} \right) \right]^{-1}, \qquad (12.2)$$

the inverse form of which yields the looked for temperature dependence of the flow stress.

Let us describe in more detail the important particular case of the negligible small viscosity $\eta \rightarrow 0$. Since $S_0$ and $S_c$ are in this case constant, $S_0 \approx 31$, $S_c \approx 27$, equation (12.2) may be easily inverted to

$$\delta\sigma = \delta\sigma_0/[1 - 0.13\exp(-6.7(T_c/T - 1))], \qquad (12.3)$$

where $\delta\sigma_0=(\hbar\beta\ln(\dot{\varepsilon}_0/\dot{\varepsilon})/31(\kappa\rho)^{1/2})$ is $\delta\sigma$ at $T=0$.

Another experimentally measured quantity is the strain rate sensitivity $\partial\sigma/\partial\ln\dot{\varepsilon}$, for which we get, differentiating equation (12.3),

$$\frac{\partial\sigma}{\partial\ln\dot{\varepsilon}} = \frac{\delta\sigma_0}{\ln(\dot{\varepsilon}_0/\dot{\varepsilon})} \frac{1-(1-0.87(1-T_c/4T))\exp[-6.7(T_c/T-1)]}{\{1-0.13\exp[-6.7(T_c/T-1)]\}^2}. \quad (12.4)$$

An employment of equation (12.3) to the description of experimental data on the low temperature mechanical tests of KCl (Suzuki and Koizumi 1985) is shown in Fig.12. The athermic anomaly of the flow stress may be described by equation (12.3) with parameters $\delta\sigma_0=4.75$ MPa, $T_c=5$K.

Another example, for LiH (Kataoka and Yamada 1985), is shown in Figures 13, 14. The parameters are $\delta\sigma_0=8$ MPa, $T_c=6.1$K, $\ln(\dot{\varepsilon}_0/\dot{\varepsilon})=22$. One may see from Figures 12-14 that the present model satisfactory reproduces the qualitative behavior of the low temperature anomalies of the experimental data for KCl and LiH.

The next example, for another class of materials, metals, is shown in Fig. 15. Contrary to the isolating materials such as KCl and LiH, in which the phonon viscosity strongly decreases with the lowering of temperature, in metals there is the electronic contribution to viscosity, which is temperature independent and retains a considerable value in the temperature range of interest. Therefore, one may expect the considerable influence of the viscosity on the flow stress in the quantum range, and this circumstance should be taken into account in a quantitative description of experiments. The data for Sn were described in detail with the total determination of all parameters in paper (Natsik *et al*. 1996), and we restrict ourselves only by one illustration. According to (Natsik *et al*. 1996), the dimensionless viscosity $\eta/(\rho(\beta b \sigma_P)^{1/2})^{1/2}=0.38$. This value was used for calculation of the theoretical temperature dependence of the flow stress in Fig. 15 in the quantum range for the normal state of the metal. Open circles in Fig. 15 show the temperature dependence of the flow stress in the superconducting state of Sn. The change of the flow stress may be qualitatively explained by the change of the electronic component of the viscosity at the NS transition. This effect is more pronounced than it is predicted by the classical pre-exponential dependence of the dislocation mobility on viscosity, however, it is in line with the direct effect of the viscosity (11.2) on the exponent of the dislocation mobility at the quantum mechanism of the motion.

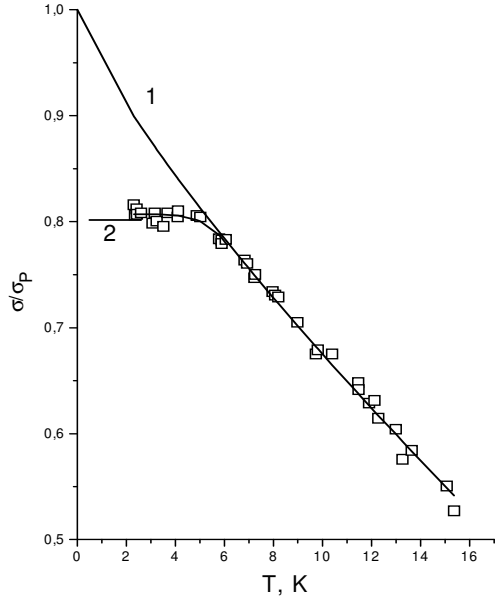

Fig. 13. The temperature dependence of the flow stress for LiH. Squares represent data (Kataoka and Yamada 1985).The line 1 shows the classical dependence (4.3), the line 2 represents the result of calculations taking into account the quantum tunneling of dislocations.

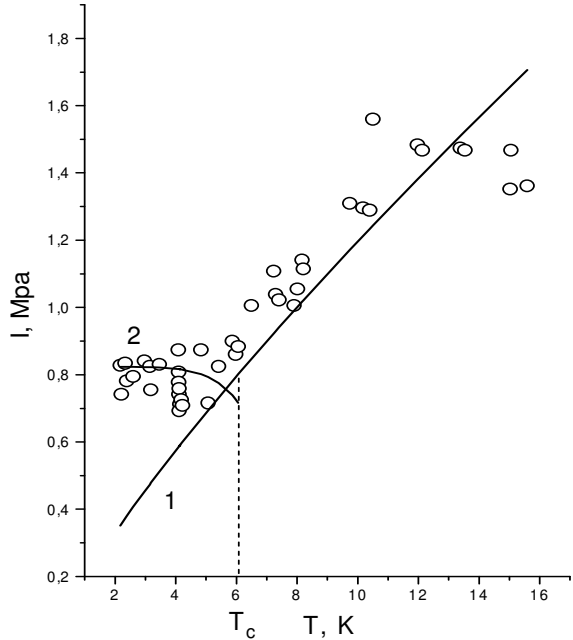

Fig. 14. The temperature dependence of the strain rate sensitivity for LiF. Circles represent data (Kataoka and Yamada 1985).The line 1 shows the classical dependence (4.3), the line 2 represents the result of calculations taking into account the quantum tunneling of dislocations. $T_c$ is the boundary between the classical and quantum temperature ranges.

Thus, the quantum mechanism of the dislocation motion provides a satisfactory explanation of the observed athermal anomalies of the plasticity of many materials. By the suitable choice of material parameters such as, for example, linear mass of the dislocation, $\rho$, the observed temperature dependencies can be described quantitatively with the formulae of the theory. However, so "weighted" dislocations prove to be significally "lighter" than it is usually expected from the estimation $\rho \sim m/b$, where $m$ is the mass of an atom of the material. For example, in LiH $\rho \approx m/8b$, for typical metals the difference is even more. The reason for such the discrepancy is unclear yet. The manifestation of the quantum effects in the thermally activated dislocation motion<, as described in Section 6, provides an incomplete explanation. It is possible that the estimation of the linear mass $\rho \sim m/b$ is not quite correct and requires an improvement. Other possible explanation is the presence of some mechanisms, which make the quantum temperature range wider, for example, an influence of impurities.

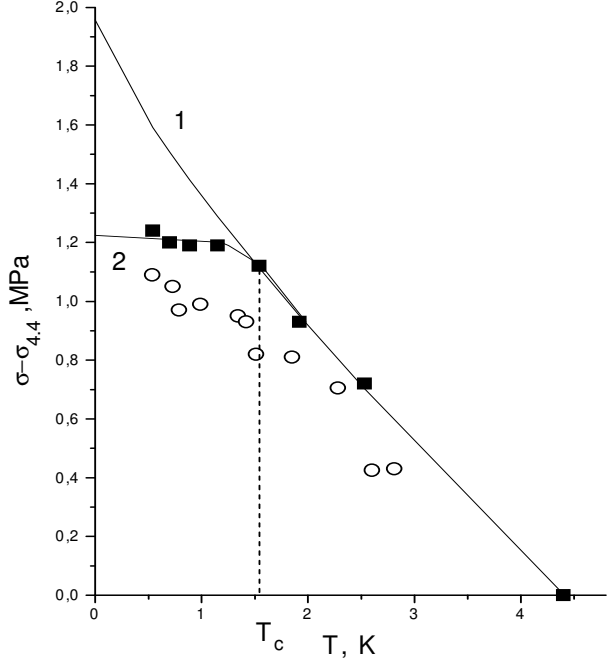

Fig. 15. The temperature dependence of the flow stress for Sn. Squares represent data (Natsik et al. 1996).The line 1 shows the classical dependence (4.3), the line 2 represents the result of calculations taking into account the quantum tunneling of dislocations. Circles show the flow stress in superconducting state. $T_c$ is the boundary between the classical and the quantum temperature ranges.

## 13. EXTENSION OF TEMPERATURE RANGE OF MANIFESTATION OF QUANTUM EFFECTS DUE TO INFLUENCE OF LOCAL OBSTACLES ON TUNNEL MECHANISM OF THE KINK PAIR FORMATION

In section 5 the influence of local obstacles, limiting the area for the thermally activated kink pair formation, was considered. Of course, the space limitation modifies also the rate of the dislocation tunnelling through the Peierls-Nabarro barrier. The aim of this Section is to investigate an influence of the local obstacles on the competition between classical thermally activated and quantum tunnel mechanism of the dislocation motion.

We shall calculate the rate of the kink pair formation at $T=0$ in framework of the same variational approach with the separation of the time and space variables by the using of the multiplicative trial function $y(x,t)=\lambda(t)q(x)$. The functional form of $\lambda(t)$ is for the infinite time interval, similarly to (10.8), $1/\cosh^2(Bt)$. We repeat now the procedure of the reducing the problem to the effectively one-dimensional one, as before, see equations (10.8), (10.9) changing this time the order of the treatment of the time- and space-dependent functions. Removing the time coordinate by the integration the action over $t$, we arrive to the problem of the minimization of the functional

$$S = \int_{-l/2}^{l/2} [\frac{B_1}{2}(\frac{dq}{dx})^2 + \frac{B_2}{2}q^2 - \frac{B_3}{3}]dx \qquad (13.1)$$

with $B_1=4/(3B)$, $B_2=16B/15+8/(3B)$, $B_3=16/(15B)$.

The solution of this problem is given in Appendix and is shown in Fig. 16. Let us consider now the limiting case of short intervals. Using the result of Appendix, equation (A.9), and minimizing it over $B_3$, we obtain

$$S \approx 10^3/l^4. \qquad (13.2)$$

It is interesting to note that the action $S$ increases with the decreasing of the interval length slower that the activation energy (5.1) does. This means that the transition temperature $T_c$ between the classical thermally activated mechanism and the tunnel mechanism of the kink pair formation is higher in the case of short intervals. For $l$ shorter than the kink pair size $\kappa^{1/2}/(\beta b \sigma_P \delta)^{1/4}$, the transition temperature to the quantum regime of the dislocation motion may be higher than in the pure material, containing no local obstacles. Qualitatively similar behavior was observed experimentally in Sn doped by Cd at small concentration of the foreign atoms of order 0.1% (Natsik *et al*. 1996).

## 14. CONCLUSION

One may conclude that the theory reviewed provides en effective tool for describing the temperature and stress dependencies of the dislocation mobility in the potential relief of the crystal lattice for both classical thermally activated and quantum mechanical tunnelling mechanisms. The theory is a semiquantitative since it is based on phenomenological Peierls-Nabarro potentials, and the value of the Peierls- Nabarro stress characterizing the amplitude of the crystal lattice relief as well as an attributed to dislocations mass must be taken from separate microscopic calculations or from experiments. Nevertheless, the predictive ability of the theory is high, since being supplemented with a few parameters, it describes the temperature dependencies of the dislocation mobility and the plastic deformation of crystals in a wide range.

Thus the presented theory realizes a compromise between the completeness of the description, on one hand, and simplicity and generality, on other hand. The comparison of the calculated dependencies with experimentally measured ones in the total available temperature range may serve for the identification of underlying mechanisms and determination of microscopic parameters of materials.

It is worth to note that the presented theory exploiting the model of an elastic string overcoming a periodic potential by fluctuations has a number of important applications in addition to that of describing plastic deformation. The model has much more general physical meaning and describes a wide class of very different phenomena, such as the dynamics of spin chains, the motion of steps on crystal faces, of domain boundaries in two-dimensional phases, of magnetic vertexes, the conductivity of quasi-one-dimensional systems, etc.

## 15. APPENDIX

*Solution of a variational problem for a limited interval*

The problem is to find an extremum of the functional

$$J\{y(x)\}= \int_{-l/2}^{l/2} [\frac{B_1}{2}(\frac{dy}{dx})^2 + \frac{B_2}{2}y^2 - \frac{B_3}{3}y^3]dy \tag{A.1}$$

calculated for functions satisfying the boundary condition $y(\pm l/2)=0$. The Euler-Lagrange equation for the functional (A.1) is

$$B_1\frac{d^2y}{dx^2} = B_2 y - B_3 y^2. \tag{A.2}$$

Its first integral, similarly to (6), has the form

$$\frac{B_1}{2}(\frac{dy}{dx})^2 - \frac{B_2}{2}y^2 + \frac{B_3}{3}y^3 = C = const. \tag{A.3}$$

Due to the symmetry of the problem in respect to the inversion $z \to -z$, the optimal function $y_0(z)$ has an extreme value (denote it $y_M$) at point $z=0$, $y(0)=y_M$. We have, therefore, $dy_0(0)/dz=0$, and it follows from (A.3) the relationship $C=-B_2 y_M^2/2 + B_3 y_M^3/3$.

The shape of the optimal function $y_0(z)$ can be found from equation (A.3) as

$$z = \pm \sqrt{\frac{B_1}{2}} \int_{y_0}^{y_M} \frac{dy}{\sqrt{\frac{B_2}{2}y^2 - \frac{B_3}{3}y^3 + C}}. \tag{A.4}$$

Using the boundary condition, we have from (A.4) equation for the determination of the integration constant $C$

$$\frac{l}{2} = \sqrt{\frac{B_1}{2}} \int_{o}^{y_M} \frac{dy}{\sqrt{\frac{B_2}{2}y^2 - \frac{B_3}{3}y^3 + C}}. \tag{A.5}$$

Substituting the solution for $y_0(z)$ (A.4) into (A.1), we arrive finally to the optimal value $J_0$ of the functional

$$J_0 = \sqrt{8B_1} \int_0^{y_M} dy \sqrt{\frac{B_2}{2}y^2 - \frac{B_3}{3}y^3 + C} - Cl. \tag{A.6}$$

Equations (A.5), (A.6) provide a parametric representation of the solution of the problem, $y_M$ is the parameter. It is illustrated in Fig. 6 for parameters $B_1=B_2=B_3=1$. This choice corresponds to the length dependence of the activation energy of the kink pair formation at stresses close to the Peierls stress (in dimensionless units).

Let us consider particular limiting cases.

*1. Long interval $l \to \infty$.*

In this case, which is analogous to the determination of the kink pair shape with the potential (2.12), $C \to 0$, integral in (A.4) can be calculated explicitly, and we have, similarly to equation (2.13),

$$y_0(x) = \frac{3B_3}{2B_3} \frac{1}{\cosh^2[\frac{z}{2}(\frac{B_2}{B_1})^{1/2}]}, \tag{A.7}$$

and

$$J_0 = \frac{6}{5} \frac{B_1^{1/2} B_2^{5/2}}{B_3^2}. \tag{A.8}$$

In particular, for $B_1=\kappa$, $B_2=2[\beta b(\sigma_P-\sigma)]^{1/2}$, $B_3=\beta$ this result coincides with (2.14).

*2. Short interval $l \to 0$.*

In this case $y_M$ increases, and the cubic term in the "potential" $B_2 y^2/2 - B_3 y^3/3$ prevails. Expanding equation (A.5), (A.6) over the small quadratic term, we arrive to the result

$$J_0 \approx \frac{72}{5} \frac{B_1^3}{B_3^2} \frac{A^6}{l^5}[1 + \frac{5}{6}\frac{B_2}{B_1}\frac{l^2}{A^3}]. \tag{A.9}$$

Here $A = \int_0^1 \frac{dx}{\sqrt{1-x^3}} \approx 1.4022...$.

## 16. LIST OF SYMBOLS

$A_1 = 0.25(5/6)^{4/5}(\beta^{3/5}k^{4/5}/b\kappa^{2/5})\ln^{4/5}(\dot{\varepsilon}_0/\dot{\varepsilon})$.
$A(T) = E(T)/kT + S(E(T))/\hbar$

| | |
|---|---|
| $a$ | lattice parameter |
| $a_2$ | period of second kind Peierls-Nabarro relief |
| $B$ | parameter characterizing space scale of optimal function |
| $b$ | Burgers vector of dislocation |
| $d$ | kink width |
| $D_k$ | kink diffusivity |
| $D_0 \sim \nu_D a_2^2$ | |
| $E(T)$ | preactivation energy at temperature $T$ |
| $E\{y(x)\}$ | multidimensional potential in space of dislocation configurations |
| $E(\sigma, l)$ | energy of kink pair formation in restricted interval |
| $E_0$ | kink pair formation energy |
| $E_k$ | kink energy |
| $E_m$ | height of kink migration barrier |
| $F = \sigma ab$ | driving force for kink |
| $g$ | kink pair formation rate per unit of dislocation length |
| $\hbar$ | Planck constant |
| $J$ | parameter (2.10) characterizing density of eigenvalues |
| $k$ | Boltzman constant |
| $L$ | length of dislocation segment |
| $l_0 = \kappa^{1/2}[\beta b(\sigma_P-\sigma)]^{-1/4}$ | kink pair size at stresses close to $\sigma_P$ |
| $m_k$ | kink mass |
| $q=(\kappa\rho)^{1/2}a^2/2\pi^2\hbar$ | |
| $q(x)$ | trial function |
| $S_0(E)$ | action at energy $E$ |
| $S(0)$ | action at zero temperature |
| $S(\sigma)$ | action for total stress range |
| $S_0\approx 31$ | |
| $\bar{S}$ | dimensionless action |
| $T$ | temperature |
| $T_c$ | temperature of transition to dislocation tunnelling |
| $T_{c2}$ | temperature of transition to kink tunnelling in secondary relief |
| $t(E)$ | period of motion over quasiclassical trajectory with energy $E$ |
| $t_{tr}$ | transition time |
| $U_0(y)$ | Peierls-Nabarro potential |
| $U(y)=U_0(y)-\sigma aby$ | inclined ("wash-board") potential |
| $U_2(x)$ | second kind Peierls-Nabarro potential relief |
| $V$ | dislocation velocity |
| $v_k$ | kink velocity |
| $x$ | coordinate along bottom of Peierls-Nabarro relief valley |
| $y$ | dislocation displacement in slip plane |
| $y(x)$ | nonuniform dislocation configuration |
| $y_0(x)$ | saddle configuration |
| $y_m$ | minimum of potential $U(x)$) |
| $y_M$ | maximum of dislocation displacement in saddle configuration. |
| $\delta y(x)$ | deviations from saddle configuration |
| $y_\alpha$ | projections of $\delta y(x)$ to main directions of potential relief at saddle point |
| $y_0$ | pass over saddle point |
| $Z_0$ | partition function for fluctuations around initial state in front of barrier |
| $Z_M$ | partition function for fluctuations around the top of the barrier |
| $Z_\alpha$ | partition function for $y_\alpha$ coordinate |
| $\beta=\vert d^3U(y_P)/dy^3\vert/2$ | |
| $\Gamma$ | kink pair generation rate |
| $\gamma=-dE(\sigma,l)/d\sigma$ | activation volume |
| $\delta=(\sigma_P-\sigma)/\sigma_P$ | |
| $\delta\sigma=\sigma_P-\sigma$ | |
| $\dot\varepsilon$ | strain rate |
| $\dot\varepsilon_0$ | pre-exponential factor in strain rate |
| $\eta$ | viscosity per unit length of dislocation |
| $k$ | line tension |
| $\lambda(t)$ | trial function |
| $\mu$ | shear modulus |
| $\mu_1$ | curvature of one-dimensional potential |
| $\mu_N=[\beta b(\sigma_P-\sigma)]^{1/2}$ | |
| $\mu_\alpha$ | eigenvalues of quadratic form of energy relief |
| $\nu_D$ | Debye frequency |
| $\Theta$ | Debye temperature |
| $\rho$ | mass per unit length of dislocation |
| $\rho(\mu)$ | density of eigenvalues |
| $\rho_d$ | density of mobile dislocations |
| $\sigma$ | external stress |
| $\sigma(T)$ | flow stress |
| $\sigma_P$ | Peierls stress |
| $\sigma_q = (\pi E_k^2/\hbar ab)(\rho/\kappa)^{1/2}$ | |
| $\phi_\alpha$ | eigenvectors of quadratic form of energy relief |
| $\Omega = (|\mu_0|/\rho + \eta^2/4\rho^2)^{1/2} - \eta/2\rho$ | |